\documentclass[usenatbib]{mnras}
\usepackage{graphicx}
\usepackage{amsmath}
\usepackage{breqn}
\usepackage{amssymb}
\usepackage{hyperref}

\usepackage{xcolor}

\title{Impact of gas-based seeding on supermassive black hole populations at $z\geq7$}
\author[Bhowmick et al.]{Aklant K. Bhowmick$^{1}$,
Laura Blecha$^{1}$,
Paul Torrey$^{2}$,
Luke Zoltan Kelley$^{3}$, \newauthor
Mark Vogelsberger$^{4}$,
Kaitlyn Kosciw$^{1}$,
Dylan Nelson$^{5}$,
Rainer Weinberger$^{6}$,
Lars Hernquist$^{6}$
\\
$^{1}$Dept. of Physics, University of Florida, Gainesville, FL 32611, USA\\
$^{2}$Dept. of Astronomy, University of Florida, Gainesville, FL, 32611, USA\\
$^{3}$Dept. of Physics and Astronomy, Northwestern University, Evanston, IL 60208, United States\\
$^{4}$Dept. of Physics, Kavli Institute for Astrophysics and Space Research, Massachusetts Institute of Technology,
Cambridge, MA 02139, USA \\
$^{5}$Universit\"{a}t Heidelberg, Zentrum f\"{u}r Astronomie, Institut f\"{u}r theoretische Astrophysik, Albert-Ueberle-Str. 2, 69120 Heidelberg, Germany\\
$^{6}$Harvard-Smithsonian Center for Astrophysics, 60 Garden Street, Cambridge, MA 02138, USA\\
}

\begin{document}
\maketitle
\begin{abstract}
Deciphering the formation of supermassive black holes~(SMBHs) is a key science goal for upcoming observational facilities. In many theoretical channels proposed so far, the seed formation depends crucially on local gas conditions. We systematically characterize the impact of a range of gas-based black hole seeding prescriptions on SMBH populations using cosmological simulations. Seeds of mass $M_{\mathrm{seed}}\sim 10^3-10^{6}~M_{\odot}/h$ are placed in halos that exceed critical thresholds for star-forming, metal-poor gas mass and halo mass (defined as $\tilde{M}_{\mathrm{sf,mp}}$ and $\tilde{M}_{\mathrm{h}}$, respectively, in units of $M_{\mathrm{seed}}$). We quantify the impact of these parameters on the properties of $z\geq7$ SMBHs. Lower seed masses produce higher black hole merger rates (by factors of $\sim10$ and $\sim1000$ at $z\sim7$ and $z\sim15$, respectively). For fixed seed mass, we find that $\tilde{M}_{\mathrm{h}}$ has the strongest impact on the black hole population at high redshift ($z\gtrsim15$, where a factor of 10 increase in $\tilde{M}_{\mathrm{h}}$ suppresses merger rates by $\gtrsim 100$). At lower redshift ($z\lesssim15$), we find that $\tilde{M}_{\mathrm{sf,mp}}$ has a larger impact on the black hole population. Increasing $\tilde{M}_{\mathrm{sf,mp}}$ from $5-150$ suppresses the merger rates by factors of $\sim8$ at $z\sim7-15$. This suggests that the seeding criteria explored here could leave distinct imprints on LISA merger rates. In contrast, AGN luminosity functions are much less sensitive to seeding criteria, varying by factors $\lesssim2-3$ within our models. Such variations will be challenging to probe even with future sensitive instruments such as Lynx or JWST. Our study provides a useful benchmark for development of seed models for large-volume cosmological simulations.
\end{abstract}


\section{Introduction}
Supermassive black holes~(SMBHs) are now believed to reside in almost every massive galaxy in the Universe~\citep{1992ApJ...393..559K,1994ApJ...435L..35H,1995Natur.373..127M}. The scaling relations between the mass of the local SMBH and their host galaxy properties \citep[e.g., stellar mass or velocity dispersion;][]{1998AJ....115.2285M,2000ApJ...539L...9F,2000ApJ...539L..13G,2013ARA&A..51..511K,2017MNRAS.471.2187D} suggest that their evolution is closely connected to the evolution of the host galaxies; these relations have been found to persist up to $z\sim1.5$~\citep{2020ApJ...896..159D}. Determining the redshifts at which these scaling relations were first established is an open problem, and is crucial to understanding SMBH formation and their subsequent evolution in relation to their host galaxies. That being said, recent simulations do suggest that these scaling relations may persist at redshifts as high as  $z\sim10$~\citep{2018MNRAS.478.5063H}, which is not very long after the epoch of seed formation. This hints that ``seeds" of SMBHs may influence the assembly of the earliest galaxies in our Universe. However, little is known about the formation mechanisms of these black hole seeds; this a major outstanding problem in astrophysics.

This is accordingly one of the key science goals of the next generation observational facilities targeting the high redshift~($z\sim6-15$) universe, 
including the James Webb Space Telescope 
\citep[JWST;][]{2006SSRv..123..485G}, the Nancy Graham Roman Space telescope 
(NGRST, formerly WFIRST; \citealt{2015arXiv150303757S}), the Lynx X-ray Observatory 
\citep{2018arXiv180909642T}, and the Laser Interferometer Space Antenna  
\citep[LISA;][]{2019arXiv190706482B}. 
Additionally, signatures of low to intermediate mass BHs~($\lesssim 10^{6}~M_{\odot}/h$ are being identified in an increasing number of $z\sim0$ dwarf galaxies, see \citealt{2013ApJ...775..116R,2017A&A...602A..28M,2019ApJ...887..245S}); this hints at the possibility that imprints of seed formation may also be present in these galaxies.      

Various channels of seed formation have been proposed so far. Remnants of Population III~(Pop III) stars~\citep{2001ApJ...550..372F,2001ApJ...551L..27M,2013ApJ...773...83X,2018MNRAS.480.3762S} may form ``light" seeds with masses of $\sim100~M_{\odot}$. Intermediate mass seeds~($\sim10^2-10^3~M_{\odot}$) could 
be formed via runaway collisions of stars in dense nuclear star clusters or NSCs~\citep{2011ApJ...740L..42D,2014MNRAS.442.3616L,2020MNRAS.498.5652K,2021MNRAS.503.1051D,2021MNRAS.tmp.1381D}. Massive seeds ($\sim10^5-10^6~M_{\odot}$) could 
be formed in atomic cooling haloes via direct collapse~(DCBHs) of pristine, low angular momentum gas in the presence of Lyman Werner~(LW) UV radiation that can prevent $H_2$ cooling \citep{2003ApJ...596...34B,2006MNRAS.370..289B,2014ApJ...795..137R,2016MNRAS.458..233L,2018MNRAS.476.3523L,2019Natur.566...85W,2020MNRAS.492.4917L}. Lastly, primordial black hole seeds~($10^2-10^5~M_{\odot}/h$) from the early universe may also contribute to SMBH populations~\citep{2001JETP...92..921R,2002astro.ph..2505K,2005APh....23..265K,2012PhLB..711....1K,2018JCAP...05..017B,2019EPJC...79..246B}.   

The discovery of a population of high redshift~($z\gtrsim6$) quasars~(with masses of $10^8-10^9~M_{\odot}$;~\citealt{2001AJ....122.2833F,2011Natur.474..616M,2015Natur.518..512W,2018Natur.553..473B}) poses a significant challenge to all these scenarios. This challenge has only become more severe with recent discoveries of massive quasars at $z\geq7$, including the current record-holder at $z=7.642$, a mere $\sim700$ Myr after the Big Bang~\citep{2021ApJ...907L...1W}. For example, low mass~($\sim10^2~M_{\odot}$) Pop III seeds require sustained Eddington or super-Eddington accretion to grow by seven 
orders of magnitude and form a $\sim10^9~M_{\odot}/h$ quasar by $z\sim6-7$. This is difficult to achieve even under optimal growth 
conditions such as a steady supply of gas and inefficient radiative feedback~(e.g., \citealt{2015MNRAS.452.1922P}). On the other hand, DCBH seeds may have a better chance of growing into the first quasars as they may form with higher initial masses in regions with significantly higher gas densities. However, the conditions required for their formation are difficult to meet because of the necessary presence of UV radiation from nearby star-forming galaxies, 
which can contaminate the region with metals before any direct collapse can occur. In a nutshell, all of these proposed seed models present challenges for producing the observed population of SMBHs. This leaves us with a large, unconstrained parameter space of possible seed models that need to be assessed for their ability to explain the observed SMBH population~(including the first quasars).  

In order to use the future observational facilities to constrain black hole seed models, it is crucial to identify their imprints on the observable properties of SMBHs and their host galaxies. For example, LISA will be able to measure merger rates at redshifts $z\gtrsim10$ for black hole masses a little below $\sim10^4~M_{\odot}/h$~\citep{2017arXiv170200786A}; we therefore need to characterize the theoretical predictions for these merger rates from different models in order to properly interpret the LISA measurements. To that end, 
many studies have been performed 
to date using semi-analytic models 
\citep[SAMs; e.g.,][]{2007MNRAS.377.1711S,Volonteri_2009, 2012MNRAS.423.2533B,2018MNRAS.476..407V, 2018MNRAS.481.3278R, 2019MNRAS.486.2336D, 2020MNRAS.491.4973D}. SAMs are efficient in terms of speed and memory usage, enabling such 
studies to systematically explore a wide range of seed model parameters. 
Overall, SAMs predict 
strong signatures of black hole seed models at high redshifts~($z\gtrsim7$), particularly for low mass 
BH merger rates and, to a lesser extent, 
the faint end 
of the luminosity function of active galactic nuclei (AGN). As an example, \cite{2018MNRAS.481.3278R} showed that light seeds produce a significantly higher number of low luminosity~($\lesssim10^{42}~\mathrm{ergs/s}$) AGN and low mass~($\lesssim10^{4}~M_{\odot}$) black hole mergers compared to heavy seeds at $z\gtrsim6$. That being said, an inevitable limitation of SAMs is that they are unable to accurately capture the internal structure and dynamics of galaxies. The detailed local environment determines the gas inflows that supply fuel for black hole seed formation and accretion, and it can therefore have a significant impact on the black hole masses and luminosities. These effects can only be captured in a hydrodynamical simulation.

Cosmological hydrodynamical simulations have emerged as powerful tools to study galaxy formation and SMBH growth~\citep{2012ApJ...745L..29D,2015MNRAS.452..575S,2015MNRAS.450.1349K,2015MNRAS.446..521S,2016MNRAS.460.2979V,2016MNRAS.463.3948D,2017MNRAS.467.4739K,2017MNRAS.470.1121T,2019ComAC...6....2N,2020MNRAS.498.2219V}. They provide us with the unique ability to understand how the different components of our contemporary galaxy formation models~(star formation, enrichment, feedback, SMBH growth) interact with the complex gas dynamics~(see also \citealt{2020NatRP...2...42V} for  recent review of the current status of cosmological simulations). The fact that black hole growth is profoundly impacted by the surrounding gas~(and vice versa) makes simulations an indispensable tool for exploring the impact of black hole seed models. Because they have a substantially higher computational demand compared to SAMs, modeling black hole seed formation in cosmological hydrodynamic simulations presents a unique set of challenges. A key limitation is the mass and spctial resolution. For most state-of-the-art large-volume cosmological simulations, densest gas cells in a halo are typically converted to black hole seeds based on a prescribed set of seeding conditions; this naturally implies that the minimum value for the seed mass is limited by the gas mass resolution which is typically $\sim10^5-10^6~M_{\odot}/h$. At the same time, a lower mass resolution of dark matter and star particles~(relative to the black hole seed mass) can also lead to dynamical instabilities~\citep{2015MNRAS.451.1868T}; this is typically dealt with by repositioning the black holes to the local potential minima in their immediate vicinity. Smaller-volume simulations including zoom-in simulations of individual halos can achieve substantially higher resolution, but this comes at the expense of statistical power and cosmic variance, especially for rare objects such as DCBHs.

Moreover, even high-resolution simulations must still rely on sub-grid models for unresolved physical processes such as star formation and chemical enrichment.
Due to all these complexities, most current cosmological simulations do not model black hole seeds based on gas properties. Instead, they seed BHs~(of mass $\sim10^5-10^6~M_{\odot}/h$) based on a threshold halo mass~(typically $\sim10^{10}~M_{\odot}/h$). These simple halo mass based prescriptions have been largely successful in reproducing properties of local SMBHs~\citep{2015MNRAS.450.1349K,2018MNRAS.474.2594M,2018MNRAS.479.4056W}. Additionally, the largest volume~($400~\mathrm{Mpc}/h$ comoving box length) cosmological hydrodynamic simulations such as  \texttt{BlueTides}~\citep{2016MNRAS.455.2778F} have also been able to reproduce the $z\gtrsim7$ quasars using a similar halo based seed model~\citep{2018MNRAS.481.4877N,2019MNRAS.483.1388T,2020MNRAS.499.3819M}. However, these models are not able to distinguish between different seed formation channels. Therefore, the next natural step is to develop realistic gas-based black hole seeding prescriptions~(representing the different seed formation channels) for the next generation of cosmological simulations. 

Efforts to incorporate gas-based seeding prescriptions have begun in the last few years. For example, the \texttt{ROMULUS} simulation~\citep{2017MNRAS.470.1121T} produces $10^6~M_{\odot}/h$ seed black holes from gas cells that have low metal fractions~($Z\lesssim10^{-4}~Z_{\odot}$), density higher than 15 times the star formation threshold, and temperatures of $\sim10^4$ K. The \texttt{Horizon-AGN} simulation~\citep{2016MNRAS.460.2979V,2017MNRAS.467.4739K} seeds $10^5~M_{\odot}/h$ black holes in regions with gas and stellar densities higher than a chosen threshold, and stellar velocity dispersion $>100~\mathrm{km/s}$. While there are several such recent works implementing physically motivated seeding prescriptions~(other examples include \citealt{2011ApJ...742...13B,2013MNRAS.428.2885D,2014MNRAS.442.2304H,2015MNRAS.448.1835T,2017MNRAS.468.3935H,2019MNRAS.486.2827D,2020arXiv200601039L,2020arXiv200204045T}), the vast parameter space of models consistent with PopIII, NSC and DCBH channels have yet to be systematically explored at a similar level as that of SAMs. To that end, the recent study of \cite{2020arXiv201201458Z} does systematically explore seed masses from $10$ to $10^6~M_{\odot}$. However, their work is a somewhat broader exploration of a variety of black hole models~(including accretion and feedback) to reproduce the most luminous $z>6$ quasars, instead of explicitly focusing on gas-based seeding and its impact on the wider SMBH population. Additionally, they had to use sub-resolution recipes for the treatment of seed masses smaller than their gas mass resolution~($6.5\times10^5~M_{\odot}/h$). Given all these developments, the time is ripe for a systematic exploration of gas-based seed models at resolutions high enough to fully resolve black hole seeds. In this study, we utilize zoom-in cosmological hydrodynamic simulations to resolve seed masses down to $\sim 10^3~M_{\odot}$.

This paper is first in a series of works dedicated towards conducting a systematic parameter study of the impact of gas-based seeding prescriptions on the high redshift~($z\gtrsim7$) black hole population. We explore the parameter space of seed models using zoom simulations. black holes are seeded in sufficiently massive halos with pristine star forming gas, which is a common feature amongst Pop III, NSC and DCBH channels. We also emphasize that our numerical seed models are completely agnostic about which seed parameters correspond to which physical channel. Different parts of the parameter space may be more~(or less) representative of different physical channels. Our approach has a unique advantage of being able to traverse the parameter space to explore many of the possible channels in the process~(note that primordial black hole seeds are not within the scope of our models). We quantify the impact of the various seed parameters on the key black hole observables such as the stellar mass vs. black hole mass~($M_*-M_{bh}$) relation, black hole mass functions, luminosity functions and merger rates. 

The structure of the paper is as follows: Section \ref{Methods} describes our basic methodology, and Section \ref{Simulation Suite} describes our simulation suite.  Section \ref{Comparison with observational constraints} compares our predictions with existing observational constraints. Section \ref{Impact of seed parameters on the BH population} focuses on the impact of our seed models on $z\gtrsim7$ SMBH populations. Section \ref{Resolution convergence} investigates the resolution convergence. Finally, we summarize our key findings in Section \ref{Summary and Conclusions}.

\section{Methods}

\label{Methods}
Our simulations were run using the \texttt{AREPO}~\citep{2010MNRAS.401..791S,2011MNRAS.418.1392P,2016MNRAS.462.2603P,2020ApJS..248...32W}  moving-mesh magneto-hydrodynamics (MHD) code.  
The code solves for gravity coupled with MHD. The gravity solver uses the PM-tree method~\citep{1986Natur.324..446B} and the MHD solver uses a non-static unstructured grid formed by performing a Voronoi tesselation of the domain. \texttt{AREPO} has been used to produce simulations of the Universe at a wide range of scales. At the largest scales, we have uniform volume cosmological simulations such as the Illustris~\citep{2014MNRAS.444.1518V,2014MNRAS.445..175G,2015A&C....13...12N,2015MNRAS.452..575S} and IllustrisTNG~\citep{2018MNRAS.475..648P,2018MNRAS.475..624N,2018MNRAS.480.5113M,2018MNRAS.477.1206N,2018MNRAS.475..676S,2019MNRAS.490.3196P,2019ComAC...6....2N,2019MNRAS.490.3234N} suites. These simulations have box sizes ranging from $\sim50$~$\mathrm{Mpc}$ to $\sim300$~$\mathrm{Mpc}$ and baryonic mass resolutions ranging from $\sim10^5-10^7~M_{\odot}$. They have been largely successful in producing galaxy and SMBH populations consistent with observations, in Illustris~\citep{2014Natur.509..177V,2015MNRAS.447L...6S,2015MNRAS.452..575S} and in TNG~\citep{2018MNRAS.479.4056W,2018MNRAS.475..648P,2018MNRAS.474.3976G,2019MNRAS.485.4817D,2019MNRAS.484.5587T,2019MNRAS.483.4140R,2021MNRAS.500.4597U,2021MNRAS.503.1940H}. At the smallest scales, we have cosmological zoom simulation suites such as AURIGA~\citep{2017MNRAS.467..179G} for individual milky-type galaxies, and  HESTIA~(High-resolutions Environmental Simulations of The Immediate Area)~\citep{2020MNRAS.498.2968L} for the Local Group. These simulations have been successful in reproducing observational results for the internal structures of galaxies~\citep{2020MNRAS.494.4291C,2020MNRAS.491.1800B,2020MNRAS.497.1603G}. All these developments make \texttt{AREPO} an ideal tool for the development of black hole models, which require a reliable modeling of the necessary physics over a large dynamic range.

In addition to gravity and MHD, \texttt{AREPO} is equipped with a wide range of ``sub-resolution" physics models. In this work, we adopt the fiducial IllustrisTNG galaxy formation model~\citep{2017MNRAS.465.3291W, 2018MNRAS.473.4077P} as our baseline model, with the obvious exception of modified black hole seed models~(described in the next section). Here we summarize some of the key 
features of our modeling that do not deviate from the IllustrisTNG model:
\begin{itemize}
\item Multiphase star forming gas is modeled using an sub-grid pressurization prescription ~\citep{2003MNRAS.339..289S}. When the densities of the gas cells exceed a threshold of $0.13~\mathrm{cm}^{-3}$, star particles are stochastically formed with an associated free-fall time scale of $2.2~\mathrm{Gyr}$.
\item Stellar evolution is implemented as described in \cite{2013MNRAS.436.3031V}, with updates for TNG as described in \cite{2018MNRAS.473.4077P}. Briefly, each star particle is assumed to represent a single stellar population~(SSP) with a fixed formation time and metallicity, with an initial mass function~(IMF) adopted from \cite{2003PASP..115..763C}. Stars return mass and metals to the nearby ISM gas following SNIa, SNII, and AGB stars. The abundances of nine species~(H, He, C, N, O, Ne, Mg, Si, Fe) are individually tracked.
\item Metal cooling is included in the presence of a spatially uniform and ionizing UV background, along with self-shielding in dense gas~\citep{2013MNRAS.436.3031V}.
\item Stellar feedback and supernovae Type II in particular produce galactic-scale winds~\citep{2018MNRAS.473.4077P}.
\item Magnetic fields are included by initializing with a small~($\sim10^{-14}~\mathrm{comoving~Gauss}$) seed magnetic field at an arbitrary orientation. Further evolution of the magnetic field is determined by the MHD equations~\citep{2011MNRAS.418.1392P}.
\item Black hole accretion is modeled using the Eddington limited Bondi-Hoyle formula given by 
\begin{eqnarray}
\dot{M}_{\mathrm{BH}}=\mathrm{min}(\dot{M}_{\mathrm{Bondi}}, \dot{M}_{\mathrm{Edd}})\\
\dot{M}_{\mathrm{Bondi}}=\frac{4 \pi G^2 M_{\mathrm{BH}}^2 \rho}{c_s^3}\\
\dot{M}_{\mathrm{Edd}}=\frac{4\pi G M_{\mathrm{BH}} m_p}{\epsilon_r \sigma_T}c
\label{bondi_eqn}
\end{eqnarray} where $G$ is the gravitational constant, $M_{\mathrm{BH}}$ is the mass of the BH, $\rho$ is the local gas density, $c_s$ is the local sound speed of the gas, $m_p$ is the mass of the proton, $\epsilon_r$ is the radiative efficiency and $\sigma_T$ is the Thompson scattering cross section. Accreting black holes have a bolometric luminosity given by 
\begin{equation}
    L=\epsilon_r \dot{M}_{\mathrm{BH}} c^2,
    \label{bol_lum_eqn}
\end{equation}
where $\epsilon_r = 0.2$ is the radiative efficiency.

\item AGN feedback is implemented in two modes. If the Eddington ratio $\eta \equiv \dot{M}_{\mathrm{bh}}/\dot{M}_{\mathrm{edd}}$ is higher than a critical value of  $\eta_{\mathrm{crit}}=\mathrm{min}[0.002(M_{\mathrm{BH}}/10^8 M_{\odot})^2,0.1]$, the feedback is deposited in the form of thermal energy that is acquired by the neighboring gas at a rate given by $\epsilon_{f,\mathrm{high}} \epsilon_r \dot{M}_{\mathrm{BH}}c^2$ where $ ~\epsilon_{f,\mathrm{high}} \epsilon_r=0.02$. $\epsilon_{f,\mathrm{high}}$ is called the ``high accretion state" coupling efficiency. If the Eddington ratio is lower than the critical value, the feedback is deposited in the form of kinetic energy that is injected into the gas at irregular time intervals. The kinetic energy manifests as a `wind' oriented along a randomly chosen direction. The injected energy is given by $\epsilon_{f,\mathrm{low}}\dot{M}_{\mathrm{BH}}c^2$ where $\epsilon_{f,\mathrm{low}}$ is called the `low accretion state' coupling efficiency. $\epsilon_{f,\mathrm{low}}$ has a maximum value of 0.2, with smaller values at very low gas densities. For further details on the AGN feedback implementation, interested readers are encouraged to refer to \cite{2017MNRAS.465.3291W}. 

\item The small-scale dynamics of black holes cannot be resolved accurately, particularly when the black hole mass is close to the gas mass resolution of the simulation. Due to this, at each time step, the black holes are ``re-positioned" on to the potential minimum within a sphere enclosing $10^3$ neighboring gas cells~(same as \texttt{IllustrisTNG}). 
\item Two black holes are promptly merged when their separations are within a distance $R_{bh}$ of at least one of the BHs; here, the $R_{bh}$ is defined to be the minimum radius of a sphere that encloses a specified number~(weighted over a smoothing kernel) of neighboring gas cells; these are listed in Table \ref{tab:my_label} for different simulation resolutions and correspond to typical $R_{bh}$ values of $\sim8~\mathrm{kpc}/h$. Note that this ``neighbor search radius" also governs the accretion and feedback properties of the BH. In particular, the density averaged over the neighboring gas cells determines the accretion rate as specified by Eq.~(\ref{bondi_eqn}). Furthermore, energy from black hole feedback is also injected into the same set of neighboring gas cells.    
\end{itemize}

\subsection{Gas-based modeling of black 
hole seeds}
\label{Gas based modeling of Black hole seeds}
Here, we describe how the black holes are seeded in our simulations based on the gas properties inside halos. Most seed formation channels that have been proposed so far have three features in common; they occur in regions with high gas density (at or above the threshold for star formation), low metallicity, and deep gravitational potentials~\citep{2012Sci...337..544V,2012AdAst2012E..12S}.

 Motivated by these considerations, 
we assume a 3-parameter family of seed models. More specifically, a black hole of seed mass $M_{\mathrm{seed}}$ is inserted in a halo if the following two conditions are met:
\begin{enumerate}
\item The total mass of star forming, metal poor gas in the halo exceeds a critical threshold denoted by $\tilde{M}_{\mathrm{sf,mp}}$. Subscripts `sf' and `mp' denote `star forming' and `metal poor' respectively. The tilde (\textasciitilde{} 
symbol) denotes that this quantity 
is in the units of $M_{\mathrm{seed}}$.

\item The total mass of the halo exceeds a critical threshold denoted by $\tilde{M}_{\mathrm{h}}$. Here again, the tilde (\textasciitilde{} 
symbol) denotes that it is in the units of $M_{\mathrm{seed}}$.
\end{enumerate}
If the above two conditions are met, the densest gas cell~(hereafter referred to as ``parent gas cell") in the halo is converted to a seed BH. 
``Metal poor " gas cells are defined as those having metallicities less than $10^{-4}~Z_{\odot}$~(note that the primordial metallicity is assumed to be $6.29\times10^{-8}~Z_{\odot}$ as in IllustrisTNG). This value is chosen because: 1) it is within the regime of Pop III stars~($\lesssim10^{-3}~Z_{\odot}$), and 2) metal line cooling can be sufficiently suppressed at these metallicities for a direct collapse to occur~\citep{2020MNRAS.494.2851C}. With this choice, our gas-based seed models are broadly consistent with physical conditions typical of all possible seed formation channels~(PopIII, NSC and DCBH). That being said, we also note that our results are not sensitive to this metallicity threshold for values ranging between $10^{-5}-10^{-2}~Z_{\odot}$. This is because the  stellar and gas metallicities in our simulation rapidly exceeds $\sim10^{-3}~Z_{\odot}$ due to the limited time resolution of our simulation compared to the stellar evolution time scales post main sequence.  


Overall, our final model contains three key variable parameters, namely $M_{\mathrm{seed}}$, $\tilde{M}_{\mathrm{sf,mp}}$, and $\tilde{M}_{\mathrm{h}}$. Our paper is primarily focused on systematically exploring the influence of these seed parameters on the resulting SMBH populations. 
Of course, these three parameters alone do not fully represent all possible physically motivated seeding channels; rather, this work is meant to provide a systematic analysis of how the basic halo and gas properties impact the SMBH population, which will inform future models that may include additional physics~(e.g. seeding based on LW intensities and infall rates of gas towards halo centers). 

\subsubsection{Additional seeding criteria: Preventing spurious seeds and mergers}
\label{Preventing spurious seeds and mergers}
As noted earlier, black holes in our simulations are repositioned on the potential minimum of their host halos to prevent dynamical artifacts from numerical noise. Along the same lines, black hole pairs are promptly merged when their separation falls below the spatial resolution scale.  To avoid potential unphysical situations caused by this, we have imposed 
a few additional seeding conditions: 


\begin{enumerate}

\item First, there are scenarios where a halo may want to seed a new BH, but that BH’s neighbor search radius~($R_{bh}$) would immediately include another black hole. Thus, if a new black hole were seeded, it would immediately merge.  This problem is exacerbated for smaller black hole seeds; this is because they form in lower mass halos which are more abundant and generally smaller in size, increasing the likelihood for the neighbor search radii to overlap with nearby halos. Additionally, the neighbor search radii~(for fixed number of neighbors) in lower mass halos are also typically larger than in higher mass halos due to lower gas densities. We therefore prevent this by not allowing the formation of any new seeds if their parent gas cells are within the neighbor search radius of pre-existing black holes.  

\item Second, there can also be scenarios where a halo can artificially lose a BH~(via re-positioning and prompt merger) during a close encounter with a much larger halo~(and much deeper potential). If the encounter turns out to be a ``fly-by"~(and does not lead to a halo merger), then the halo will get re-seeded with a new BH. This overall leads to an artificial increase in the number of black hole seeds and black hole mergers. While we cannot prevent the halo from losing its black hole with the current black hole re-positioning scheme, we can certainly prevent the creation of new ``spurious" seeds. In order to do that, we ``tag" the neighboring gas cells of every newly seeded halo; this retains the memory of past seeding events in halos. We then compute the amount of ``tagged gas" in a halo at a given time step, and use it as a proxy to determine if the halo contains imprints of a past seeding event. We thereby prevent re-seeding from occurring in those halos. 
Appendix \ref{Suppressing short lived seeds and spurious reseeds} contains the full details of this implementation and demonstrates
that if we do not suppress these spurious seeds, the merger rates can be overestimated by up to factors of $\sim10$.      
\end{enumerate}

\section{Simulation Overview}
\label{Simulation Suite}
\subsection{Zoom simulation suite}
We apply the black hole seeding methods described in the previous subsection to a suite of cosmological zoom-in simulations targeted at helping us understanding the impact of gas-based black hole seeding. Therefore, our simulation suite mainly consists of these zoom runs 
with a parent volume of $(25~\mathrm{Mpc}/h)^3$. The cosmology is based on the results of \cite{2016A&A...594A..13P}, i.e. $\Omega_{\Lambda}=0.6911, \Omega_m=0.3089, \Omega_b=0.0486, H_0=67.74~\mathrm{km}~\mathrm{sec}^{-1}\mathrm{Mpc}^{-1},\sigma_8=0.8159,n_s=0.9667$. The simulations were initialized at $z=127$ using the \texttt{MUSIC}~\citep{2011MNRAS.415.2101H} initial condition generator. 

The resolution of the background grid and high-resolution zoom regions are characterized by parameters we refer to as $L_{\mathrm{min}}$~($\mathrm{levelmin}$) and $L_{\mathrm{max}}$~($\mathrm{levelmax}$), respectively. The resolution level, $L$, is 
defined such that a uniform-resolution box with the same mass resolution would have $2^L$ DM particles per side. 
We set $L_{\mathrm{min}}=7$ for the background grid, which corresponds to a DM mass resolution of $5.3\times10^9~M_{\odot}/h$. For the high resolution zoom region, we consider $L_{\mathrm{max}}$ values ranging from $9$ to $12$. Additionally, we also have a buffer region comprising of DM particles at intermediate resolutions between the background grid and the zoom region; this buffer region enables a smooth transition between zoom region to background grid. The details of the mass and spatial resolution for various values of $L_{\mathrm{max}}$ are shown in Table \ref{tab:my_label}. For the highest resolution runs with $L_{\mathrm{max}}=12$, the DM mass resolution is $1.6 \times 10^4~M_{\odot}/h$ and gas mass resolution is 
$\sim10^3~M_{\odot}/h$~(note that masses of gas cells depend on the extent of refinement or derefinement of the Voronoi cells). However, the $L_{\mathrm{max}}=12$ simulations have significant computational expense, and we only run a select few of them in order to 
1) resolve 
seed masses of $\sim10^3~M_{\odot}$ 
and to 2) investigate resolution convergence. We demonstrate in Section \ref{Resolution convergence} that our simulations approach resolution convergence for $L_{\mathrm{max}}\geq11$. Therefore, the majority of our seed parameter space is explored using $L_{\mathrm{max}}=11$ simulations, which have a gas mass resolution of $\sim10^4~M_{\odot}/h$.

\begin{table*}
     \centering
    \begin{tabular}{c|c|c|c|c|c|}
         $L_{\mathrm{max}}$ & $M_{dm}$~($M_{\odot}/h$) & $M_{gas}$~($M_{\odot}/h$) & $\epsilon~(kpc/h)$ & Black hole neighbors & Allowed $M_{\mathrm{seed}}$~($M_{\odot}/h$) values\\
         \hline
         9 & $8.4\times10^6$ & $\sim10^6$ & 1 & 32 & $8\times10^{5}$\\
         10 & $1\times10^6$ & $\sim10^5$ & 0.5 & 64 & $8\times10^{5}$, $1\times10^{5}$ \\
         11 & $1.3\times10^5$ & $\sim10^4$ & 0.25 & 128 & $8\times10^{5}$, $1\times10^{5}$, $1.25\times10^{4}$  \\
         12 & $1.6\times10^4$ & $\sim10^3$ & 0.125 & 256 & $8\times10^{5}$, $1\times10^{5}$, $1.25\times10^{4}$ , $1.56\times10^{3}$ \\
         \hline
    \end{tabular}
    \caption{Spatial and mass resolutions within the zoom region of our simulations for various values of $L_{\mathrm{max}}$~(see Section \ref{Simulation Suite} for the definition). $M_{dm}$ is the mass of a dark matter particle, $M_{gas}$ is the typical mass of a gas cell~(note that gas cells can refine and de-refine depending on the local density), and $\epsilon$ is the gravitational smoothing length. The 4th column represents the number of nearest gas cells that are assigned to be black hole neighbors. The 5th column corresponds to the seed masses allowed at each $L_{\mathrm{max}}$, which is limited by the gas mass resolution.}
    \label{tab:my_label}
\end{table*}

\subsubsection{The parameter space~($M_{\mathrm{seed}}$, $\tilde{M}_{\mathrm{sf,mp}}$, $\tilde{M}_{\mathrm{h}}$) of seed models}
The seed masses $M_{\mathrm{seed}}$ range from $8\times10^{5}$ to $1.56\times10^3~M_{\odot}/h$, as listed in Table \ref{tab:my_label}. Note that the minimum seed mass that can be probed at each $L_{\mathrm{max}}$ is limited by the gas mass resolution of the simulation. For example, at $L_{\mathrm{max}}=10$, we can explore only seed masses 
$\geq 1\times10^5~M_{\odot}/h$. But at the highest resolution of $L_{\mathrm{max}}=12$, we can explore seed masses 
$\geq 1.56\times10^3~M_{\odot}/h$. The differences in mass resolution between successive values of $L_{\mathrm{max}}$ is roughly a factor of 8; 
for this reason we also vary 
$M_{\mathrm{seed}}$ by multiples of 8.  

The parameter $\tilde{M}_{\mathrm{sf,mp}}$ is a multiplicative factor that~(when multiplied by $M_{\mathrm{seed}}$) determines how much of the star forming, metal poor gas mass is required to form a black hole seed of mass $M_{\mathrm{seed}}$. A larger value of $\tilde{M}_{\mathrm{sf,mp}}$ therefore corresponds to a lower efficiency for converting pristine star-forming gas into black hole seeds. This efficiency may depend on the seed formation channel as well as the halo properties~(e.g., mass, concentration, or spin), but there are currently no empirical constraints. We choose to explore values of  $\tilde{M}_{\mathrm{sf,mp}}=5,50,150$ which correspond to seed formation efficiencies of $20,2,0.7\%$, respectively.    

The parameter $\tilde{M}_{\mathrm{h}}$ is similarly a multiplicative factor that, when multiplied by $M_{\rm seed}$, sets the minimum threshold halo mass 
for the formation of black hole seeds. We explore values of $\tilde{M}_{\mathrm{h}}=10^3,~3\times10^3~\&~ 10^4$. Previous work~\citep{2003ApJ...592..645Y,2020arXiv201004169K} has shown that the critical halo mass for the collapse of molecular gas clouds is $\sim10^{5}-10^{6}~M_{\odot}$. If Pop III seeds of $10^{2}~M_{\odot}$ form in halos above these critical masses, this corresponds to $\tilde{M}_{\mathrm{h}}\sim 10^3-10^4$. At the other end, direct collapse of gas can allow formation of $\sim10^4-10^{5}~M_{\odot}$ black holes in atomic cooling halos of masses close to $\sim10^7-10^8~M_{\odot}$, again corresponding to $\tilde{M}_{\mathrm{h}}\sim 10^3-10^4$~(this is broadly based on results of \citealt{2019Natur.566...85W} and 
\citealt{2020OJAp....3E..15R}). Therefore, our choice of varying $\tilde{M}_{\mathrm{h}}$ from $10^3-10^4$ is reasonable for an initial exploration of the parameter space. That being said, we do not rule out models with $\tilde{M}_{\mathrm{h}}$ higher than $10^4$, but they would form too few black holes within our zoom volume to robustly predict the statistical properties of SMBH populations.       
\subsubsection{Target regions for high resolution zoom-ins}
\label{Target regions for high resolution zoom-ins}
\begin{figure*}
\includegraphics[width=10.5 cm]{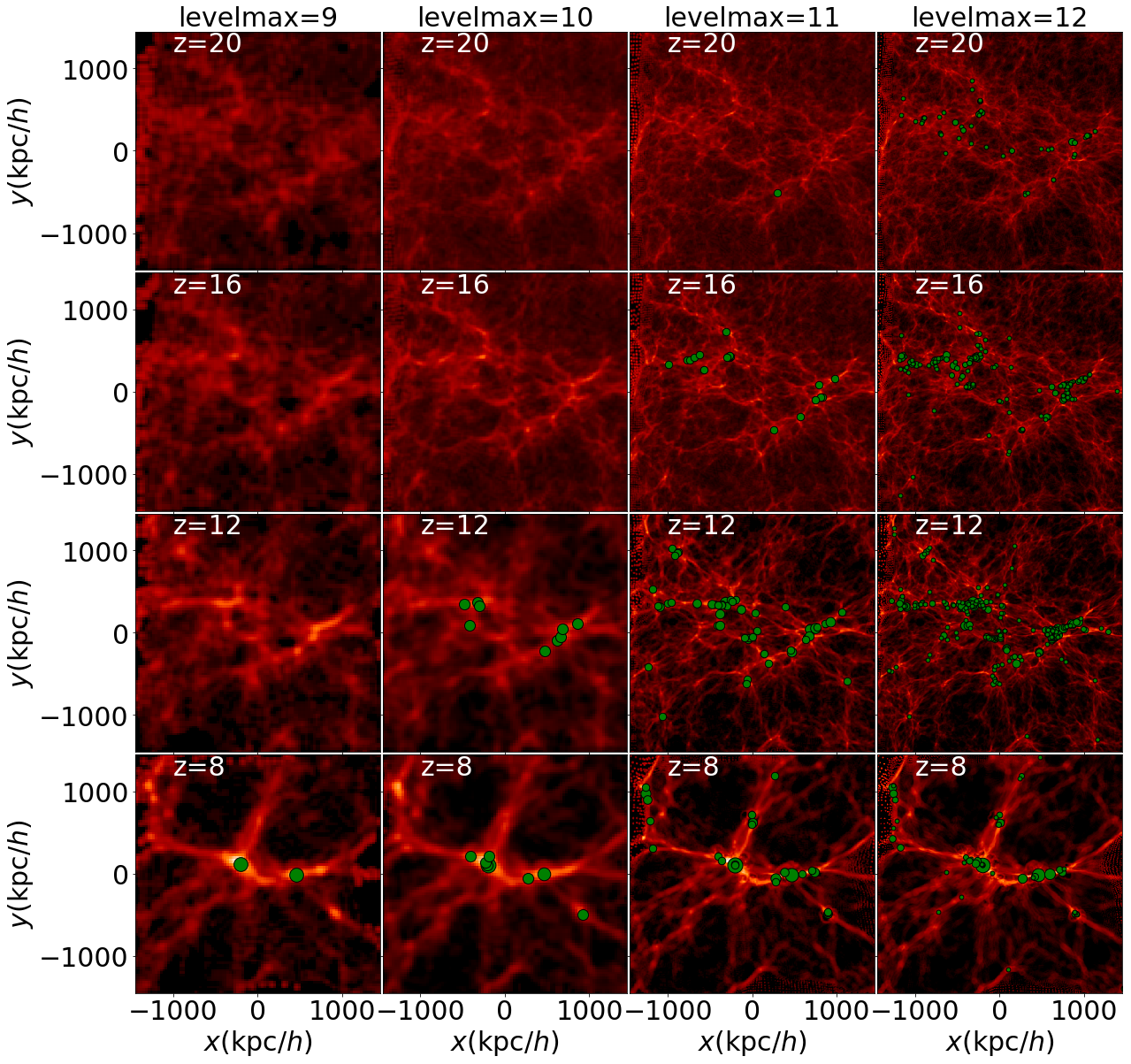}
\includegraphics[height=10.1 cm]{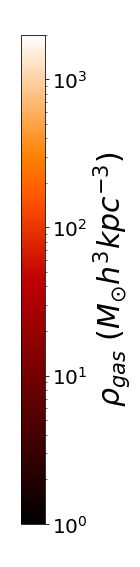}

\caption{2D color map of the gas density projection in \texttt{ZOOM_REGION_z5}~($3000\times3000\times50$ kpc/h). Left to right panels correspond to increasing resolution. Top to bottom panels show the redshift evolution from $z=20$ to $z=8$. The green circles represent BHs~(bigger circles are more massive BHs). Here, we show simulations 
where the seed masses are close to gas mass resolution. Therefore $L_{\mathrm{max}}=9,10,11,12$ correspond to $M_{\mathrm{seed}}=8\times10^{5},1\times10^5,1.25\times10^4,1.56\times10^3~M_{\odot}/h$ respectively. As the resolution increases, the simulations reveal finer structures that 
are able to capture the formation of lower-mass seed black holes at increasingly earlier epochs.}
\label{density_2dplot_fig}

\includegraphics[width=10.5 cm]{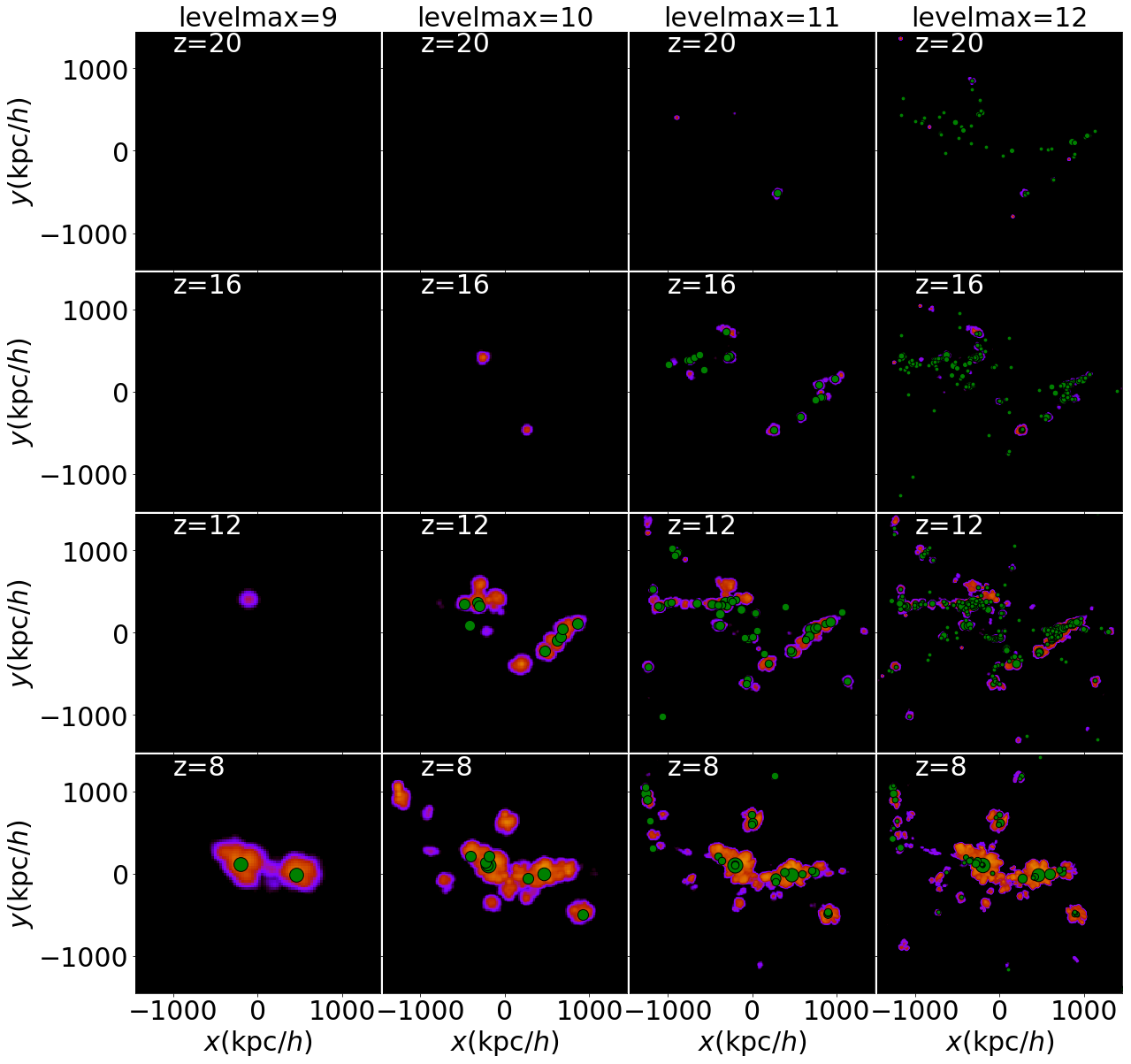}
\includegraphics[height=10.1 cm]{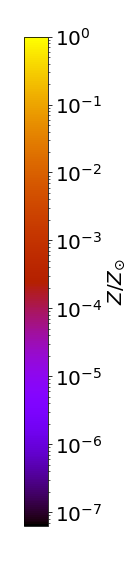}


\caption{2D color map of the gas metallicity projection in \texttt{ZOOM_REGION_z5}~($3000\times3000\times50$ kpc/h), for the same simulation snapshots as in Figure \ref{density_2dplot_fig}. As the resolution increases, 
the onset of star formation and metal enrichment happen at earlier epochs. However, the star formation and metal enrichment begin to approach convergence at $\mathrm{L_{\mathrm{max}}\geq11}$. 
}
\label{metallicity_2dplot_fig}
\end{figure*}

Our suite is comprised of two target regions. We first run a low resolution uniform volume simulation with 25 $\mathrm{Mpc}/h$ box-size and $128^3$ particles~($L_{\mathrm{min}}=L_{\mathrm{max}}=7$) from $z=127$ to $z=0$. Our parent box-size is large enough to produce target regions that are sufficiently overdense to allow formation of a significant number of seeds. Larger box-sizes would allow us to simulate more extreme regions, but at significantly higher computational expense. We select two target regions from our uniform box as follows:
\begin{itemize}
    \item Our primary target region is focused on making predictions for the high redshift universe~($z\gtrsim7$). We select a target halo of mass $3.5\times10^{11}~M_{\odot}/h$ at $z=5$~(corresponding to a peak height\footnote{The peak height $\nu$ of a halo quantifies the ``extremeness" of the corresponding region in terms of its overdensity. It is defined as $\nu\equiv\delta_c/\sigma(R,z)$ where $\sigma^2(R,z)$ is the variance of the overdensity in a sphere of lagrangian radius $R$ of the halo, specified by a spherical top hat window function as the weighing kernel. $\delta_c=1.686$ is the critical overdensity for a spherically symmetric collapse in an Einstein De-Sitter universe. Please refer to \cite{2008ApJ...688..709T} for more details.}  of $\nu=3.3$) and trace its DM particles to $z=127$. We then construct a cubical volume with minimum dimensions required to enclose the particles, and assign it to be the initial zoom volume at $z=127$ (one could alternatively choose an ellipsoid or convex hull, but this would not impact our main results). We shall hereafter refer to this region as $\texttt{ZOOM_REGION_z5}$. Figures \ref{density_2dplot_fig} and \ref{metallicity_2dplot_fig} show the density and metallicity profiles for $\texttt{ZOOM_REGION_z5}$ at various redshift snapshots between $z=20$ to $z=8$. As the density increases, star formation ensues and that is soon followed by formation of seed black holes. The subsequent stellar evolution processes lead to the onset of metal enrichment; the metals continue to disperse throughout the region and eventually suppresses the formation of new black holes at $z\sim11-12$~(this is demonstrated more quantitatively in Section \ref{Resolution convergence}). The number of seeds that form in this region~(up to $z\sim7$) range from $\sim5-1000$ depending on the seed mass; models with lower $M_{\mathrm{seed}},\tilde{M}_{\mathrm{h}}~\&~\tilde{M}_{\mathrm{sf,mp}}$ naturally lead to a higher number of seeds~(we shall quantify these trends in Section \ref{Seed formation times}).

    \item Our secondary target region is focused on making predictions at $z=0$ and comparing against observational constraints. We follow the same procedure as $\texttt{ZOOM_REGION_z5}$, but for a target halo of mass $6.4\times10^{11}~M_{\odot}/h$ at $z=0$~(this corresponds to a peak height of $\nu=0.8$). We choose a lower density region~(compared to \texttt{ZOOM_REGION_z5}) to reduce the computational expense of running simulations to $z=0$. We shall hereafter refer to this region as $\texttt{ZOOM_REGION_z0}$. The number of seeds formed in $\texttt{ZOOM_REGION_z0}$ is $\sim50\%$ fewer than that of $\texttt{ZOOM_REGION_z5}$.     
\end{itemize}

The peripheral regions of the zoom volume are inevitably susceptible to contamination from low resolution DM particles. Note that our halo mass criterion for seeding is based on the mass contributed exclusively by the highest resolution DM particles; as a result, most~($\gtrsim95\%$) of our black holes seed in regions which have negligible levels~($\lesssim1\%$) of contamination from low resolution DM particles. However, as the region undergoes gravitational collapse, some black holes inevitably end up in halos which are contaminated by low resolution particles. When the host halos~(BH environments) are contaminated, the accretion rates~(and therefore also their masses and luminosities) are susceptible to numerical artifacts. We find that within $1~\mathrm{Mpc}/h$ from the center of mass~(COM) of the zoom volume, there is virtually no contamination at all redshifts above $z\gtrsim7$. Therefore, when analyzing the final black hole masses and luminosities at various snapshots in Sections \ref{BH mass functions} and \ref{Luminosity Functions}, we include only those BHs~(at each snapshot) which are within $1~\mathrm{Mpc}/h$ from the center of mass of the zoom volume. By doing this, we exclude $\sim70\%$ of all black holes at $z\sim7$~(the percentage decreases at higher redshifts); while this is a significant amount, the remaining black holes are assured to have masses and luminosities that are not impacted by numerical artifacts. Note however that for analysing the number of seeding and merger rates in Sections \ref{Seed formation times} and \ref{Merger rates} respectively, we include all the events that occur within our zoom volume. The seeding frequency is not significantly impacted by low resolution contamination because most seeds do start out in regions with negligible contamination. As for the merger rates, due to our black hole re-positioning scheme, their ``dynamics" is largely unaffected by the low resolution contamination; therefore, the merger rates are also not impacted by the presence of low resolution DM particles. We verified these by re-running simulations where the seeding is additionally restricted to halos with a maximum low resolution contamination $<1\%$; we find that this additional criterion does not significantly impact the rates of seeding and merger events. 

\subsubsection{Nomenclature for the zoom simulation runs}
\label{Nomenclature for the zoom simulation runs}
Here we briefly describe the nomenclature we use to hereafter~(in Figure legends and captions) identify the different boxes amongst the large suite of simulations described throughout Section \ref{Simulation Suite}. 

The general structure of our label is represented as ``\texttt{L*_SM*_FOF*}" wherein each ``\texttt{*}" following ``\texttt{L}", ``\texttt{SM}" and ``\texttt{FOF}" is replaced by the numerical value of 
$L_{\mathrm{max}}$, $\tilde{M}_{\mathrm{sf,mp}}$ and $\tilde{M}_{\mathrm{h}}$ respectively. The value of $M_{\mathrm{seed}}$ is explicitly stated along with the label. As an example, the zoom run with $L_{\mathrm{max}}=10$, $M_{\mathrm{seed}}=1\times10^{5}~M_{\odot}/h$, $\tilde{M}_{\mathrm{sf,mp}}=50$, and $\tilde{M}_{\mathrm{h}}=3\times10^3$ is labelled as ``\texttt{L10_SM50_FOF3000} with $M_{\mathrm{seed}}=1\times10^{5}~M_{\odot}/h$".

\subsection{Uniform volume simulation suite}
\label{Uniform volume simulation suite}
In order to validate the gas-based seed models, our first step is to compare our seed model predictions with observations. To reliably compare with the volume-independent constraints of the luminosity functions and black hole mass functions, our predictions should have sufficiently low cosmic variance. Our zoom simulation predictions inevitably have a large cosmic variance due to their small volumes. Furthermore, the shape of the zoom volume becomes complex at late times due to its gravitational collapse, further complicating efforts to obtain volume-independent predictions from zoom simulations. For these reasons, we run a set of three uniform volume simulations with $25~\mathrm{Mpc}/h$ box-size and $512^{3}$ particles~(using the same initial condition). The models for these uniform boxes are described as follows:

\begin{itemize}
\item The first and second boxes use the gas-based seed models~(see Section \ref{Gas based modeling of Black hole seeds}). Both the boxes have  $M_{\mathrm{seed}}=8\times10^{5}~M_{\odot}/h$ and $\tilde{M}_{\mathrm{h}}=3\times10^3$. $\tilde{M}_{\mathrm{sf,mp}}$ is chosen to be $5$~(referred as \texttt{UNIFORM_SM5}) and $50$~(referred as  \texttt{UNIFORM_SM50}) for the first and second box respectively.
\item The third box~(referred as \texttt{UNIFORM_FOF}) uses the default halo mass based seed model from the Illustris-TNG suite. More specifically, seeds of $M_{\mathrm{seed}}=8\times10^{5}~M_{\odot}/h$ are placed in halos above a mass of $5\times10^{10}~M_{\odot}/h$. Hereafter, we shall also refer to this as the `TNG model'. 
\end{itemize}

\section{Comparison with observational constraints}
\label{Comparison with observational constraints}
\begin{figure*}
\includegraphics[width=7cm]{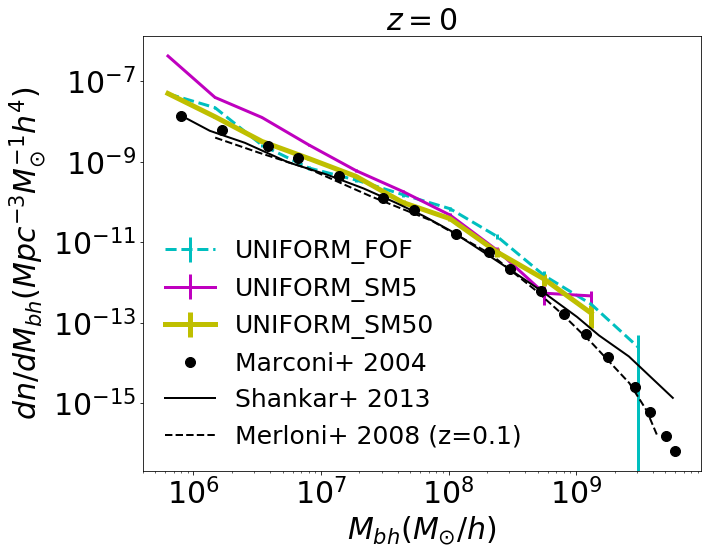}  \includegraphics[width=6.75cm]{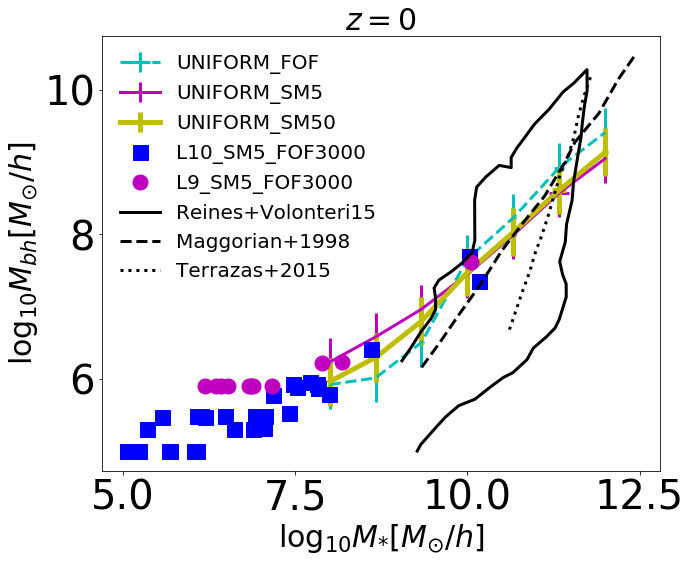} 

\caption{BH mass functions~(left panel) and $M_*-M_{bh}$~(right panel) relations at $z=0$. In the right panels, $M_*$ and $M_{bh}$ correspond to the total stellar mass and black hole mass of subhalos~(we only include subhalos with $>10$ particles; also true for subsequent figures). The solid pink and light green lines correspond to uniform volume simulations using gas-based seed models~($\tilde{M}_{h}=3\times10^3$) with $\tilde{M}_{\mathrm{sf,mp}}=5$~(\texttt{UNIFORM_SM5}) and $50$~(\texttt{UNIFORM_SM50}) respectively with $M_{\mathrm{seed}}=8\times10^5~M_{\odot}/h$. In the right panel, we also have pink and blue squares which are predictions from zoom simulations~(\texttt{ZOOM_REGION_z0} with $\tilde{M}_{h}=3\times10^3$, $\tilde{M}_{\mathrm{sf,mp}}=5$) with $L_{\mathrm{max}}=9$~(\texttt{L9_SM5_FOF3000} with $M_{\mathrm{seed}}=8\times10^5~M_{\odot}/h$) and 10 (\texttt{L10_SM5_FOF3000} with $M_{\mathrm{seed}}=1\times10^5~M_{\odot}/h$) respectively. The dashed cyan line corresponds to the default halo mass based seed model~(\texttt{UNIFORM_FOF}) used in \texttt{IllustrisTNG}. The black points and solid line~(left panel) are observational measurements of the local black hole mass function from \protect\cite{2004MNRAS.351..169M} and  \protect\cite{Shankar_2013} respectively; these are also consistent with other local measurments \protect\citep[for e.g.][]{2004MNRAS.354.1020S,2005SSRv..116..523F,2007ApJ...663...53T,Shankar_2013,Mutlu_Pakdil_2016} for black holes $\gtrsim10^7~M_{\odot}$~(see Fig. 5 of  \protect\citealt{2009ApJ...690...20S} and Fig. 13 of \protect\citealt{Mutlu_Pakdil_2016}). The black dashed line in the left panel is the constraint from \protect\cite{2008MNRAS.388.1011M} at $z=0.1$. The black solid line~(right panel) corresponds to the approximate outline for the observational scatter from \protect\cite{2015ApJ...813...82R}. Black dashed and dotted lines~(right panel) are mean stellar bulge mass vs. black hole mass relations from \protect\cite{1998AJ....115.2285M} and \protect\cite{2016ApJ...830L..12T} respectively. The error-bars in the left panel are Poisson errors. The error-bars in the right panel are obtained via bootstrap resampling. Given the observational uncertainties, the measurements are broadly consistent with the simulation predictions.}
\label{BH_mass_observations_fig}

\includegraphics[width=15.5cm]{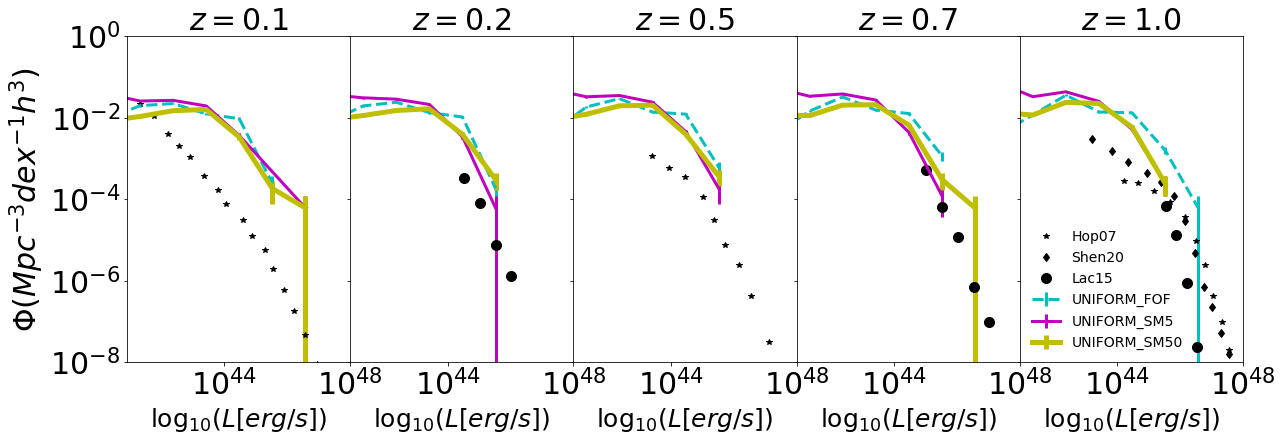} 
\includegraphics[width=15.5cm]{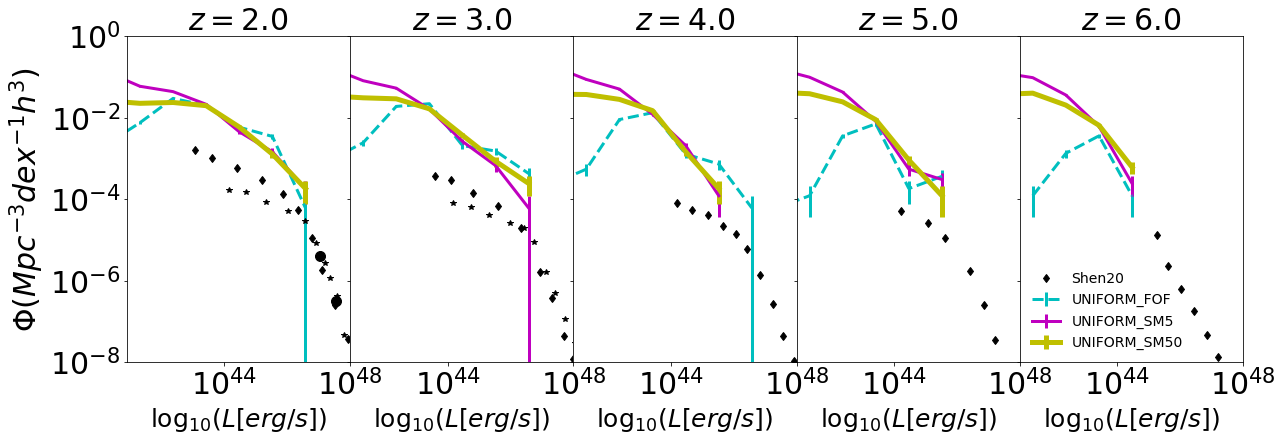} 
\caption{Luminosity functions between $0.1<z<6$ for a set of uniform volume simulations. The solid pink and light green lines correspond to gas-based seed models~($\tilde{M}_{h}=3\times10^3$) with $\tilde{M}_{\mathrm{sf,mp}}=5$~(\texttt{UNIFORM_SM5}) and $50$~(\texttt{UNIFORM_SM50}) respectively. The dashed cyan line corresponds to the default halo mass based seed model~(\texttt{UNIFORM_FOF}) used in \texttt{IllustrisTNG}. The error-bars are Poisson errors. The black points correspond to observational constraints from \protect\cite[Hop07]{2007ApJ...654..731H}, \protect\cite[Lac15]{2015ApJ...802..102L} and \protect\cite[Shen20]{2020MNRAS.495.3252S}. Across all redshifts, the luminosity functions predicted by the gas-based seed models are consistent with the TNG model, particularly in the regimes probed by both simulations and observations.}
\label{LF_observations_fig}
\end{figure*}

In this section, we use the uniform volume simulation suite~(described in Section \ref{Uniform volume simulation suite}) to compare and validate our gas-based seed models against existing observational constraints. In Figure \ref{BH_mass_observations_fig}, we compare our model predictions of black hole masses to that of the observed black holes in the local Universe. The left panel shows the local~($z=0$) black hole mass functions predicted by our uniform boxes and compares it to $z\sim0$ observational constraints~\citep{2004MNRAS.351..169M,2004MNRAS.354.1020S,2005SSRv..116..523F,2007ApJ...663...53T,2008MNRAS.388.1011M,Shankar_2013, Mutlu_Pakdil_2016}. We find that all three (\texttt{UNIFORM_SM5}, \texttt{UNIFORM_SM50}, \texttt{UNIFORM_FOF}) models produce remarkable agreement with the observed measurements at $M_{bh}\gtrsim10^{7}~M_{\odot}/h$. At $10^6\lesssim M_{bh} \lesssim 10^{7}~M_{\odot}/h$, the \texttt{UNIFORM_SM50} and \texttt{UNIFORM_FOF} models fully agree with the measurements but the \texttt{UNIFORM_SM5} model produces slightly higher number of black holes compared to the observations. However, this difference~($\sim0.7~\mathrm{dex}$) may be difficult to be distinguishable given the observational uncertainties~($\sim0.5~\mathrm{dex}$, see Figure 6 of \citealt{2004MNRAS.351..169M}). 



The right panel of Figure \ref{BH_mass_observations_fig} compares the model predictions for local $M_*-M_{bh}$ relations to measurements from \cite{1998AJ....115.2285M}, \cite{2015ApJ...813...82R} \& \cite{2016ApJ...830L..12T}. We find that the predictions for all the uniform boxes are broadly consistent with the measurements at $M_{bh}\gtrsim10^{7}~M_{\odot}/h$ living in galaxies with $M_{*}\gtrsim10^{10}~M_{\odot}/h$. For lower-mass black holes and galaxies, our models predict slightly higher black hole masses compared to the measurements. Additionally, at the highest mass end~($M_*\sim10^{12}~M_{\odot}/h$ galaxies), simulations seem to mildly underpredict the black hole masses; however, we also expect a significant amount of cosmic variance at the most massive end due to our limited box size. Note that these differences between simulations and observations are also present for the TNG model, as also pointed out in \cite{2020MNRAS.493..899H}; but it is nevertheless encouraging to see that our gas-based seed models produce local SMBH populations that are in good agreement with the TNG model. This also suggests that the local SMBH population is relatively insensitive to the seed models for black hole masses $\gtrsim10^6~M_{\odot}/h$; in this regime, differences between theory vs. observations likely point towards other aspects of the black hole models such as accretion and feedback. Notably, these results are overall consistent with \cite{2008MNRAS.383.1079V} which finds that for the local black hole population, differences between seed model predictions persist only at the low mass end~($\lesssim10^6~M_{\odot}/h$) of the black hole mass function.   
In addition to the uniform volume boxes, we also make predictions for zoom simulations for the region ``\texttt{ZOOM_REGION_z0}"~(see Section \ref{Target regions for high resolution zoom-ins}). The target galaxy in this zoom region has a mass of $\sim10^{10}~{M_\odot}/h$~(see the most massive galaxy in Figure \ref{BH_mass_observations_fig}: right panel). We ran \texttt{L9_SM5_FOF3000} with $M_{\mathrm{seed}}=8\times10^5~M_{\odot}/h$ and \texttt{L10_SM5_FOF3000} with $M_{\mathrm{seed}}=1\times10^5~M_{\odot}/h$, and find that both of them produce an SMBH of mass $3\times10^7~M_{\odot}/h$ in the target galaxy; this is also consistent with the observed measurements.

Figure \ref{LF_observations_fig} shows the bolometric luminosity functions of the uniform volumes, compared with observed luminosity functions at $z \sim 0 - 6$. 
\citep{2007ApJ...654..731H,2015ApJ...802..102L,2020MNRAS.495.3252S}. 
All of the uniform volume simulations produce similar luminosity functions, at least at the high-luminosity end. 
At high redshifts~($z\gtrsim2$), the differences between model predictions start to become distinguishable at $\lesssim10^{42}-10^{43}~\mathrm{ergs/s}$. These luminosities shall be accessible with upcoming X-ray facilities~(e.g. Lynx;~\citealt{2020arXiv201102501S}). 

Next, we focus on comparing the model predictions to observational constraints in Figure \ref{LF_observations_fig}. Note in general that the overlapping regime between the simulation predictions and the observed measurements is small. But in regimes where they do overlap, the simulation predictions are broadly consistent with the measurements of \cite{2015ApJ...802..102L} at $z\sim0.2-1$. We also find that the bright ends of the predicted luminosity functions are~(when extrapolated) consistent with the bright ends of the observations at all redshifts between $0.2-6$. However, there are some differences too between the measurements and our predictions. In particular, the measurements of \cite{2007ApJ...654..731H} and \cite{2020MNRAS.495.3252S} indicate a flattening of the slope at their faint end at $\sim10^{45}~\mathrm{ergs/s}$. But the slope predicted by the simulations does not flatten until much lower luminosities~($\sim10^{43}~\mathrm{ergs/s}$). In fact, the slopes for the simulation predictions at $\sim10^{43-45}~\mathrm{ergs/s}$ continue to be similar to those of the measurements at the brightest end~($\sim10^{45}-10^{47}~\mathrm{ergs/s}$). As a result, the simulations predict significantly higher numbers of AGN between $\sim10^{41}-10^{45}~\mathrm{ergs/s}$ compared to \cite{2007ApJ...654..731H} and \cite{2020MNRAS.495.3252S}. 

The tension between the simulation vs observed luminosity functions has also been noted earlier by \cite{2018MNRAS.479.4056W}, when comparing the predictions from IllustrisTNG. As noted in their work, there are a number of sources of uncertainty which can affect both the simulation predictions and observed measurements at the low luminosity end. For example, the simulation predictions depend sensitively on the radiative efficiency at low Eddington ratios; this may be different from the constant value of 0.2 as assumed by us~(for instance, in general relativistic MHD simulations of accretion flows in \citealt{2017MNRAS.468.1398S}, the radiative output increases with increase in Eddington ratio  from $10^{-6}$ to $10^{-2}$). At higher redshifts in particular, the discrepancies can also be potentially alleviated by a more efficient stellar feedback, as proposed by \cite{2017MNRAS.467..179G} to account for insufficient suppression of star formation in $z\sim2-3$ galaxies. At the same time, the observed 
luminosity functions are also uncertain, particularly the number of Compton thick AGN at low luminosities~\citep{2015ApJ...802...89B}. Furthermore, evidence of significant obscuration in $\lesssim10^{45}~\mathrm{ergs/s}$ AGN is also found in recent works of \cite{2019MNRAS.484.4413H} and \cite{2020arXiv201102501S}. That being said, models for \cite{2007ApJ...654..731H} and \cite{2020MNRAS.495.3252S} do account for obscuration, despite which the tension persists. Future measurements will shed further light on the tension between the simulations and observations at the low luminosity end~($\sim10^{42}-10^{45}~\mathrm{ergs/s}$) of the observed luminosity functions. But it is nevertheless encouraging to see that the different seed models have similar predictions in this regime. 


Overall, we find that our gas-based seed models produce $z\sim0$ SMBH and $z\lesssim7$ AGN populations that are fully consistent with the TNG model. As a result, they also do not severely conflict with existing observational constraints, particularly for the black hole masses at $z\sim0$. However, differences do exist~(for both gas-based as well as TNG models) between simulations and observations for the AGN luminosity functions. But these differences seem to be largely insensitive to the seed model, and therefore points to other aspects of the black hole modeling such as accretion and feedback. Exploring the space of feedback and accretion models to study their impact on the luminosity functions will be a subject of future work. Here, we shall continue to explore our gas-based seed models to systematically assess their impact on the $z\geq7$ SMBH population.   

.
\section{Impact of seed parameters on the $z\geq7$ SMBH population}
\label{Impact of seed parameters on the BH population}
We now investigate how our seed parameters $M_{\mathrm{seed}}$, $\tilde{M}_{\mathrm{sf,mp}}$ and $\tilde{M}_{\mathrm{h}}$ affect the various observable properties of our SMBH population at $z\gtrsim7$. 
\subsection{BH seed formation times}
\label{Seed formation times}
\begin{figure*}
\includegraphics[width=16cm]{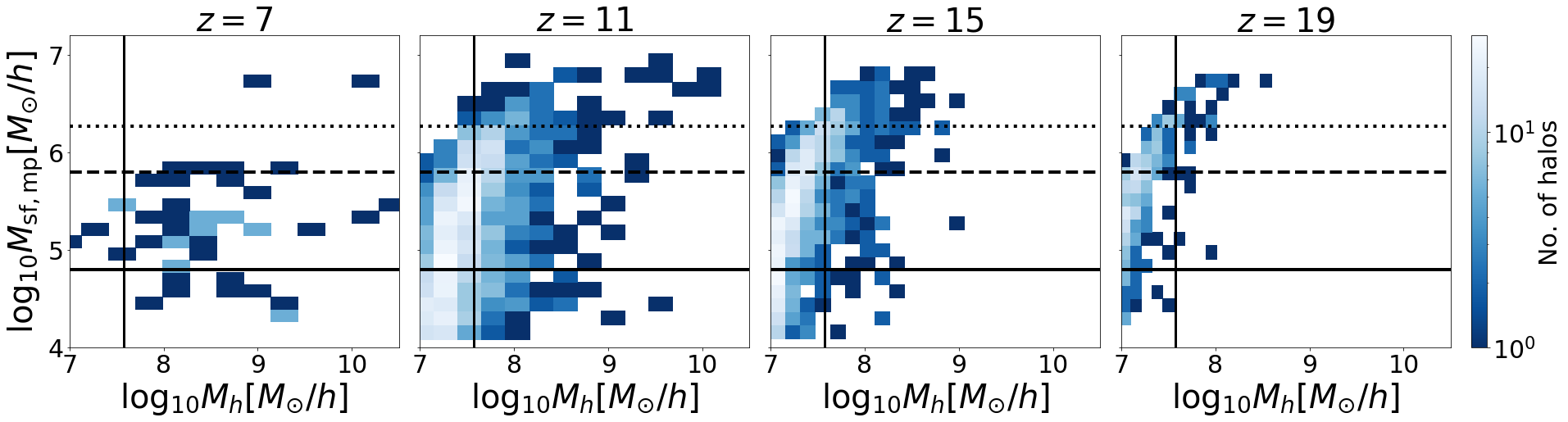} 
\caption{Blue histograms show the total halo mass~($M_h$) vs star forming, metal poor gas mass~($M_{\mathrm{sf,mp}}$) for the halo populations at $z=7-19$ within \texttt{ZOOM_REGION_z5}. The following assumes $M_{\mathrm{seed}}=1.25\times10^4~M_{\odot}/h$. The vertical solid line corresponds to the seeding threshold of $\tilde{M}_{h}=3000$ for the total halo mass. The solid, dashed and dotted horizontal lines correspond to seeding thresholds of $\tilde{M}_{\mathrm{sf,mp}}=5,50$ and $150$ for the star forming, metal poor gas mass. Generally, halos with higher total halo mass tend to have higher star forming, metal poor gas mass; however, there is significant scatter~(up to $\sim2~\mathrm{dex}$). As a result of the scatter, increasing $\tilde{M}_{\mathrm{sf,mp}}$ from $5$ to $150$ does not significantly impact the minimum total mass of halos in which seeds can form.}
\label{Halo_properties_figg}
\end{figure*}

We first look at how the rates of seed formation are impacted by changing $M_{\mathrm{seed}}$, $\tilde{M}_{\mathrm{sf,mp}}$ and $\tilde{M}_{\mathrm{h}}$. 
Before we proceed, it is important to understand the potential interdependencies between our three seed parameters. We first emphasize that the seeding frequency in our model effectively depends only on halo mass and star-forming, metal poor gas mass threshold. These thresholds are scaled to the seed mass $M_{\mathrm{seed}}$; therefore, any impact of changing $M_{\mathrm{seed}}$ is a simple consequence of the commensurate change in the total halo mass and star-forming, metal poor gas mass threshold. Figure \ref{Halo_properties_figg} shows the correlation between the total halo mass and star-forming, metal poor gas mass. Not surprisingly, halos with higher total mass typically have higher star forming, metal poor gas mass~(particularly for halos at $z\gtrsim11$ when metal enrichment is not that prevalent). However, the scatter between the total halo mass and star forming, metal poor gas mass is significant~(up to $\sim2~\mathrm{dex}$); this is large enough that increasing the star forming, metal poor gas mass threshold does not significantly impact the minimum total mass of halos that can form seeds. Additionally, at lower redshifts~($z\lesssim11$), increased metal enrichment in halos generally leads to decrease in the star forming, metal poor gas mass, whereas the total halo mass continues to increase. These considerations motivate us to independently study the impact of $\tilde{M}_{\mathrm{sf,mp}}$ and $\tilde{M}_{\mathrm{h}}$; as we shall see, varying $\tilde{M}_{\mathrm{sf,mp}}$ and $\tilde{M}_{\mathrm{h}}$ indeed leads to qualitatively distinct impacts on seed formation. 

\begin{figure*}
\includegraphics[width=6.5cm]{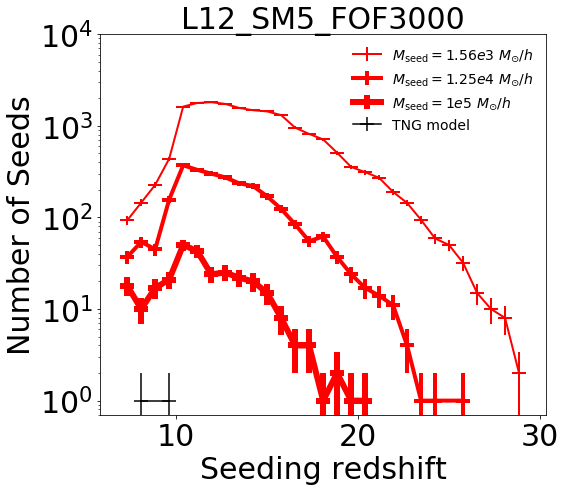} 

\caption{\textbf{Dependence of seed formation times on $M_{\mathrm{seed}}$:} Distribution of seed formation times for models with different $M_{\mathrm{seed}}$ at fixed $\tilde{M}_{\mathrm{sf,mp}}=5$ and $\tilde{M}_{h}=3\times10^3$~(for \texttt{ZOOM_REGION_z5}). The $\tilde{M}_{\mathrm{sf,mp}}$ and $\tilde{M}_{h}$ values can be inferred from the label ``\texttt{L*_SM*_FOF*}", wherein each ``\texttt{*}" following ``\texttt{L}", ``\texttt{SM}" and ``\texttt{FOF}" is replaced by the numerical value of 
$L_{\mathrm{max}}$, $\tilde{M}_{\mathrm{sf,mp}}$ and $\tilde{M}_{\mathrm{h}}$ respectively. The value of $M_{\mathrm{seed}}$ is explicitly stated along with the label. We adopt this convention to label our zoom boxes throughout the remaining figures of the paper. The error-bars in the number counts are Poisson errors~(this is true for all the remaining figures which show number counts) These simulations are all run at $\mathrm{L_{\mathrm{max}}=12}$. The distributions peak at $z\sim11$. As $M_{\mathrm{seed}}$ increases, there is a delay in the onset of seeding and concurrently, the number of seeding events is 
suppressed at all redshifts.}  
\label{Seed_distributions_seed_dependence_fig}

\includegraphics[width=12cm]{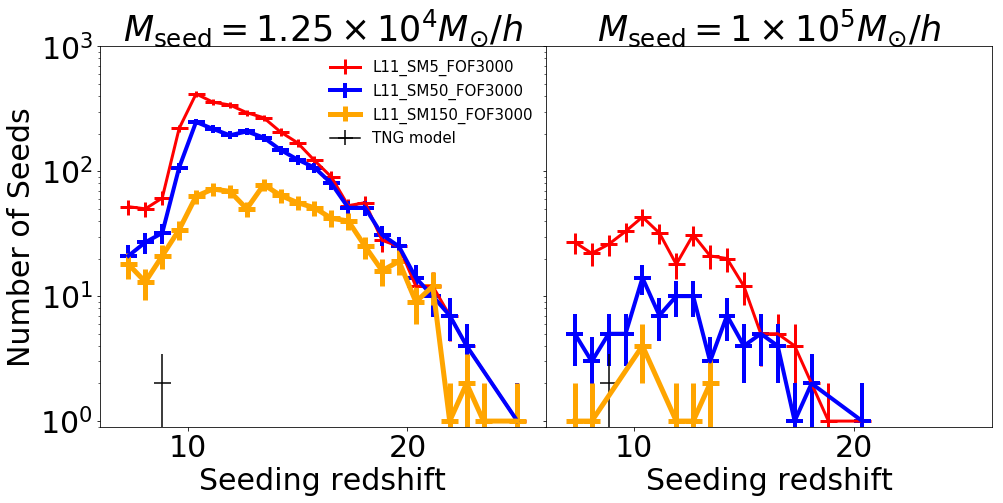} 
\caption{\textbf{Dependence of seed formation times on $\tilde{M}_{\mathrm{sf,mp}}$:} Distribution of seed formation times for models with different star-forming, metal-poor gas mass thresholds $\tilde{M}_{\mathrm{sf,mp}}$ at fixed $M_{\mathrm{seed}}$ and $\tilde{M}_{h}=3\times10^3$~(for \texttt{ZOOM_REGION_z5}). These simulations are all run at $\mathrm{L_{\mathrm{max}}=11}$. Left and right panels correspond to seed masses of $M_{\mathrm{seed}}=1.25\times10^{4}~M_{\odot}/h$ and $M_{\mathrm{seed}}=1\times10^{5}~M_{\odot}/h$, respectively. As $\tilde{M}_{\mathrm{sf,mp}}$ is increased, seeding is suppressed largely at $z\sim 7-15$,
with relatively little impact at $z\gtrsim15$. Additionally, the impact of $\tilde{M}_{\mathrm{sf,mp}}$ is stronger for higher mass seeds.}
\label{Seed_distributions_SM_dependence_fig}

\includegraphics[width=12cm]{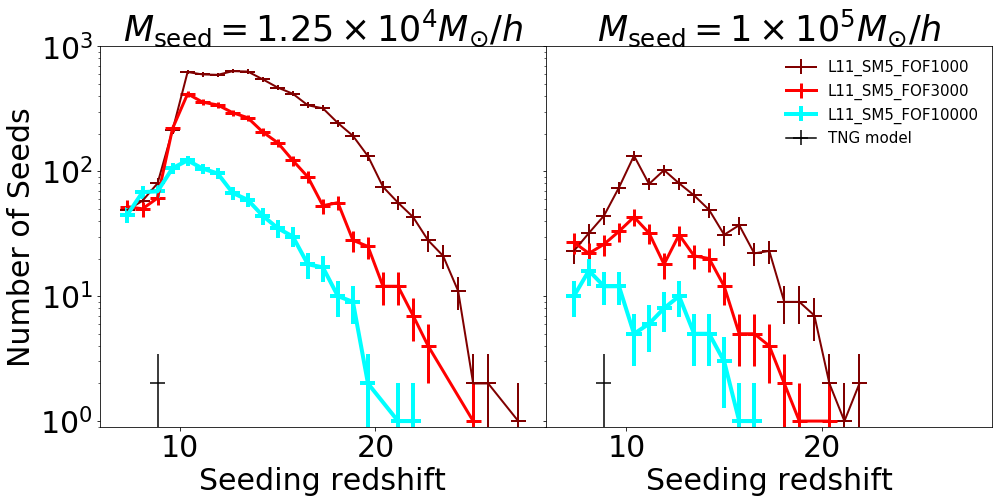} 
\caption{\textbf{Dependence of seed formation times on $\tilde{M}_{\mathrm{h}}$:} Similar to Figure \ref{Seed_distributions_SM_dependence_fig}, but for different total halo mass thresholds $\tilde{M}_{\mathrm{h}}$ at fixed $M_{\mathrm{seed}}$ and $\tilde{M}_{\mathrm{sf,mp}}=5$. The distributions peak at $z\sim11$. Increasing $\tilde{M}_{\mathrm{h}}$ uniformly suppresses the seeding throughout $z \sim 11-25$, but no significant suppression is seen at $z\sim11$. The impact of  $\tilde{M}_{\mathrm{h}}$ is similar for both seed masses.} 
\label{Seed_distributions_FOF_dependence_fig}
\end{figure*}

Figures \ref{Seed_distributions_seed_dependence_fig}, \ref{Seed_distributions_SM_dependence_fig} and \ref{Seed_distributions_FOF_dependence_fig} show the distribution of seeding redshifts for variations in $M_{\mathrm{seed}}$, $\tilde{M}_{\mathrm{sf,mp}}$ and $\tilde{M}_{\mathrm{h}}$, respectively, for \texttt{ZOOM_REGION_z5}. We find that the seeds first start to form at redshifts ranging from $z\sim15-25$ depending on the parameters. As the simulation evolves, the number of seeds continues to increase up to $z\sim11$. This is largely driven by the onset of star formation. 
In all of the gas-based seed models, the distributions of seed formation times peak at $z\sim11$; between $z\sim11\pm0.76$, we can have $10-1000$ seeds formed depending on the seed parameters. At $z\sim7-11$ however, new seed formation begins to be suppressed by contamination of star-forming regions with metals. In contrast to these gas-based seed models, the TNG model~(black lines in Figures \ref{Seed_distributions_seed_dependence_fig}-\ref{Seed_distributions_FOF_dependence_fig}) produces only two seeds by $z=7$, both of which do not form until $z\sim9$. This is due to the much higher 
halo mass threshold in the TNG model~($5\times10^{10}~M_{\odot}/h$) than in 
the gas-based models~($\sim10^6-10^{9}~M_{\odot}/h$). 

The fact that our seed formation peaks at $z\sim11$ is broadly consistent with previous works using hydrodynamic simulations such as \cite{2017MNRAS.470.1121T} and \cite{2018ApJ...861...39D} where seeding is limited to regions that are metal poor as well as star forming; note that both of these studies emulate seeding conditions for DCBHs. It is also instructive to compare with SAMs, which form a relatively larger body of the existing literature. \cite{2012MNRAS.421.1465D} studied high redshift black hole formation using seeding criteria consistent with NSC and Pop III channels. In their work, the NSC seeds were formed in star forming regions with $10^{-5}\lesssim Z\lesssim10^{-3}Z_{\odot}$; most of these seeds formed at $z\sim11$, similar to our results. However, their Pop III seeds~(which formed from Pop III stellar remnants $\gtrsim260~M_{\odot}$ and $Z\lesssim 10^{-5}~Z_{\odot}$) tend to mostly form at $z\gtrsim20$, which is significantly earlier than our model predictions. Similar results are seen for the Pop III channels in \cite{2014MNRAS.442.3616L} and \cite{2019MNRAS.483.3592B}. Overall, we find that our typical seed formation times of $z\sim11$ are in reasonable agreement with previous works that emulate DCBH and NSC seeding conditions. However, works that emulate Pop III seeding conditions tend to produce seeds  substantially earlier~($z\gtrsim20$). This is not surprising given that 1) our seed masses~($10^3-10^5~M_{\odot}/h$) are significantly higher than expected values for Pop III seeds~($\sim10^2~M_{\odot}/h$), and 2) our galaxy formation model does not include the full set of physics necessary for an accurate modelling Pop III star formation in mini-halos~(halo masses $\lesssim10^5~M_{\odot}/h$). In particular, we do not model the formation of molecular hydrogen and its associated cooling; due to this, gas is not allowed to effectively cool below temperatures $\lesssim10^4~\mathrm{K}$ in mini-halos. In future work, we shall explore galaxy formation models which do include these necessary physics, and push our simulation resolutions to fully resolve Pop III seeds. The remainder of this subsection focuses on how the overall rates of seed formation are impacted by $M_{\mathrm{seed}}$, $\tilde{M}_{\mathrm{sf,mp}}$ and $\tilde{M}_{\mathrm{h}}$.


Figure \ref{Seed_distributions_seed_dependence_fig} illustrates that, as expected, fewer seeds form when the seed mass is larger. This results directly from our choice to scale the halo and gas mass thresholds to the seed mass. At $z\gtrsim 11$, the number of seeds formed is roughly linearly proportional to the seed mass; i.e., a $\sim 100$ times larger seed mass produces $\sim 100$ times fewer seeds. 
A natural corollary to this is 
the fact that lower mass seeds start forming at earlier epochs. As we shall see, at the highest redshifts these trends are primarily driven by the impact of increasing the halo mass threshold. At slightly lower redshifts ($z\sim7-11$), the number of seeds formed in each snapshot begins to decline. In this regime, differences between $M_{\mathrm{seed}}$ models in Figure \ref{Seed_distributions_seed_dependence_fig} are instead driven primarily by the impact of increasing the star-forming, metal-poor gas mass threshold. We see less variation in the number of seeds formed; 
i.e., $M_{\mathrm{seed}}=1\times10^{5}~M_{\odot}/h$ produces $\sim10$ times fewer seeds compared to $M_{\mathrm{seed}}=1.56\times10^3~M_{\odot}/h$. 
In the next two paragraphs, we shall disentangle the impacts of star forming, metal poor gas mass threshold and halo mass threshold on the number of seeding events.

Figure \ref{Seed_distributions_SM_dependence_fig} illustrates how the threshold mass of star-forming, metal-poor gas impacts the seed formation redshifts, for  $\tilde{M}_{\mathrm{sf,mp}}=5$, 50, \& 150. Let us first focus on $M_{\mathrm{seed}}=1.25\times10^4~M_{\odot}/h$~(Figure \ref{Seed_distributions_SM_dependence_fig}: left panels). We can clearly see that higher values of $\tilde{M}_{\mathrm{sf,mp}}$ suppress seed formation, particularly at $z\lesssim15$.
The stronger suppression of seed formation at $z\lesssim15$ is largely driven by metal enrichment and dispersion~(see also Figure \ref{StarFormationMetalEnrichmentevolution_fig}). Higher redshifts mark the earliest epochs of star formation, such that almost all the star forming gas mass is still metal poor and the seeding is primarily governed by the total amount of star forming gas mass in halos~(i.e. $\tilde{M}_{\mathrm{sf,mp}}\approx \tilde{M}_{\mathrm{sf}}$). When $\tilde{M}_{\mathrm{sf,mp}}$ is scaled to a larger seed mass of $M_{\mathrm{seed}}=1\times10^5~M_{\odot}/h$, seed formation is more strongly suppressed~(right panels of Figure \ref{Seed_distributions_SM_dependence_fig}). For example, at $\tilde{M}_{\mathrm{sf,mp}}=5$, the overall number of $1\times10^5~M_{\odot}/h$ seeds are $\sim10$ times lower compared to $1.25\times10^4~M_{\odot}/h$ seeds. At higher values of $\tilde{M}_{\mathrm{sf,mp}}$~($=50,150$), even fewer seeds are formed. Here too, we see hints of stronger suppression at $z\lesssim15$~(similar to $1.25\times10^4~M_{\odot}/h$ seeds but with poorer statistics). 


Lastly, we look at the impact of $\tilde{M}_{\mathrm{h}}$, which is shown in Figure \ref{Seed_distributions_FOF_dependence_fig}. For both $M_{\mathrm{seed}}=1.25\times10^4~\&~ 1\times10^5~M_{\odot}/h$~(right $\&$ left panels respectively), we find that increasing $\tilde{M}_{\mathrm{h}}$ from $10^3$ to $10^4$ leads to nearly uniform suppression of the number of seeds by a factor $\sim10$ across the entire redshift range between $z\sim11-25$. This is a straightforward consequence of the halo mass function, which also suppresses by $\sim10$ when the threshold mass is increased from $10^3$ to $10^4~M_{\mathrm{seed}}$. Equivalently, we can also say that at $z\sim11-25$, the halo mass criterion dominates the seeding because it is much more stringent than the star forming, metal poor gas mass criterion. At $z\sim7-11$ however, the halo mass criterion has much less impact on seed formation. This is because by $z\sim11$, most halos that satisfy the star-forming, metal-poor gas mass criterion also satisfy the total halo mass criterion. 

Contrasting the impact of the halo mass threshold~($\tilde{M}_{\mathrm{h}}$) and star forming, metal poor gas mass threshold~($\tilde{M}_{\mathrm{sf,mp}}$), we see 
that the halo mass threshold uniformly impacts the seeding at $z\sim11-25$ but does not impact the seeding at $z\sim7-11$. On the other hand, the star-forming, metal-poor 
gas mass threshold impacts the seeding at $z\sim7-15$ but not at $z\sim20-25$. These distinct effects of $\tilde{M}_{\mathrm{sf,mp}}$ and $\tilde{M}_{\mathrm{h}}$ are primarily driven by the relative rates of halo growth vs. star formation vs metal enrichment. At higher redshifts, it is the lack of sufficiently massive star forming halos that is the main barrier for seed formation; therefore, $\tilde{M}_{\mathrm{h}}$ has a significant impact. As we approach lower redshifts, the growth of star forming halos does trigger seed formation; however, this is soon accompanied by metal enrichment that ultimately halts the formation of new seeds; therefore, $\tilde{M}_{\mathrm{sf,mp}}$ starts to have a more significant impact at lower redshifts. 

\begin{figure}
\includegraphics[width=8cm]{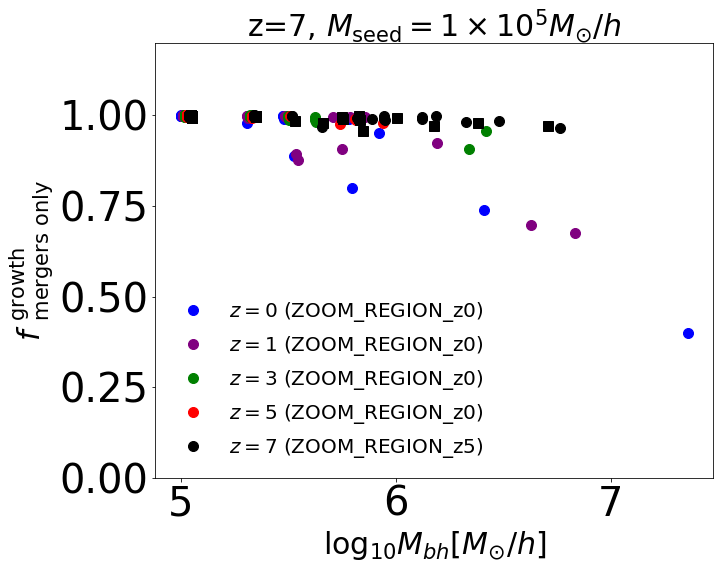} 
\caption{$f^{\mathrm{growth}}_{\mathrm{mergers~only}}$ is the fraction of mass that a black hole accumulates via mergers only. This is defined as $f^{\mathrm{growth}}_{\mathrm{mergers~only}}=(N_{\mathrm{prog}} M_{\mathrm{seed}})/M_{bh}$ where $N_{\mathrm{prog}}$ is the number of progenitors of the black hole at a given snapshot. The black points show the results for the $z=7$ snapshot of \texttt{ZOOM_REGION_z5} for models with $\tilde{M}_{\mathrm{sf,mp}}=5,50~\&~150$~(circles, squares \& stars respectively). At $z\geq7$, the growth is almost completely 
driven by black hole mergers, since the accretion rates are small. The purple, green and red cicles show the redshift evolution of $f^{\mathrm{growth}}_{\mathrm{mergers~only}}$ from $z=7$ to $z=0$ snapshot of \texttt{ZOOM_REGION_z0}~($\tilde{M}_{\mathrm{sf,mp}}=5$). This shows that as we move to lower redshifts, the gas accretion starts to become increasingly important for more massive black holes. However, mergers continue to remain the dominant contributor~($\gtrsim50\%$) to black hole growth for $M_{bh}\sim10^4-10^7~M_{\odot}/h$ even at $z=1$.}
\label{acc_vs_mergers_fig}
\end{figure}

\subsection{Growth of BHs: Mergers vs. Accretion}
\label{Growth of BHs: Mergers vs. Accretion}
Following seed formation, a black hole can grow by merging with other BHs, or by accreting the surrounding gas. Here we investigate the relative contributions of mergers vs.~accretion in fueling black hole growth; this is shown in Figure \ref{acc_vs_mergers_fig}~(black points) for black holes at the $z=7$ snapshot of \texttt{ZOOM_REGION_z5}. We find that black holes predominantly grow via merging with other black holes at $z\geq7$ for all of the parameter values we explore in this work. This is likely because: 1) low mass black holes are not able to accrete significantly within the Bondi Hoyle prescription wherein the accretion rate is $\propto M_{bh}^2$, 2) most of the black holes form in regions where the gas densities are not yet high enough to significantly fuel black hole accretion. Additionally, it is important to note that the mergers are being facilitated by our repositioning scheme. A more realistic model for black hole dynamics can lead to dynamical delays and reduce (or at least postpone) the number of mergers; this can curb the overall amount of merger driven black hole growth. Even though this paper focuses on $z\geq7$, it is instructive to see how mergers and accretion contribute to black hole growth at lower redshifts. This is also shown in Figure \ref{acc_vs_mergers_fig}~(purple, green and red points) from $z=7$ to $0$ for \texttt{ZOOM_REGION_z0}. We find that as we approach $z\sim0$, accretion starts to have an increasingly significant contribution to the black hole growth for increasingly massive black holes. However, the contribution of mergers still continues to be dominant~($\gtrsim50\%$) even at $z=1$ for black holes with masses up to $10^7~M_{\odot}/h$.   

We also note that there are alternate accretion models where the accretion rate scales differently with $M_{bh}$; for example, in the gravitational torque driven accretion models~\citep{2011MNRAS.415.1027H,2017MNRAS.464.2840A,2019MNRAS.486.2827D}, the accretion rate scales as $M_{bh}^{1/6}$. In such models, the early black hole growth at low masses would be much more rapid compared to that of Bondi accretion. This would 
likely increase the contribution of accretion to black hole growth at $z\geq7$.    

\subsection{BH masses}
\label{BH mass functions}
\begin{figure*}
\includegraphics[width=12cm]{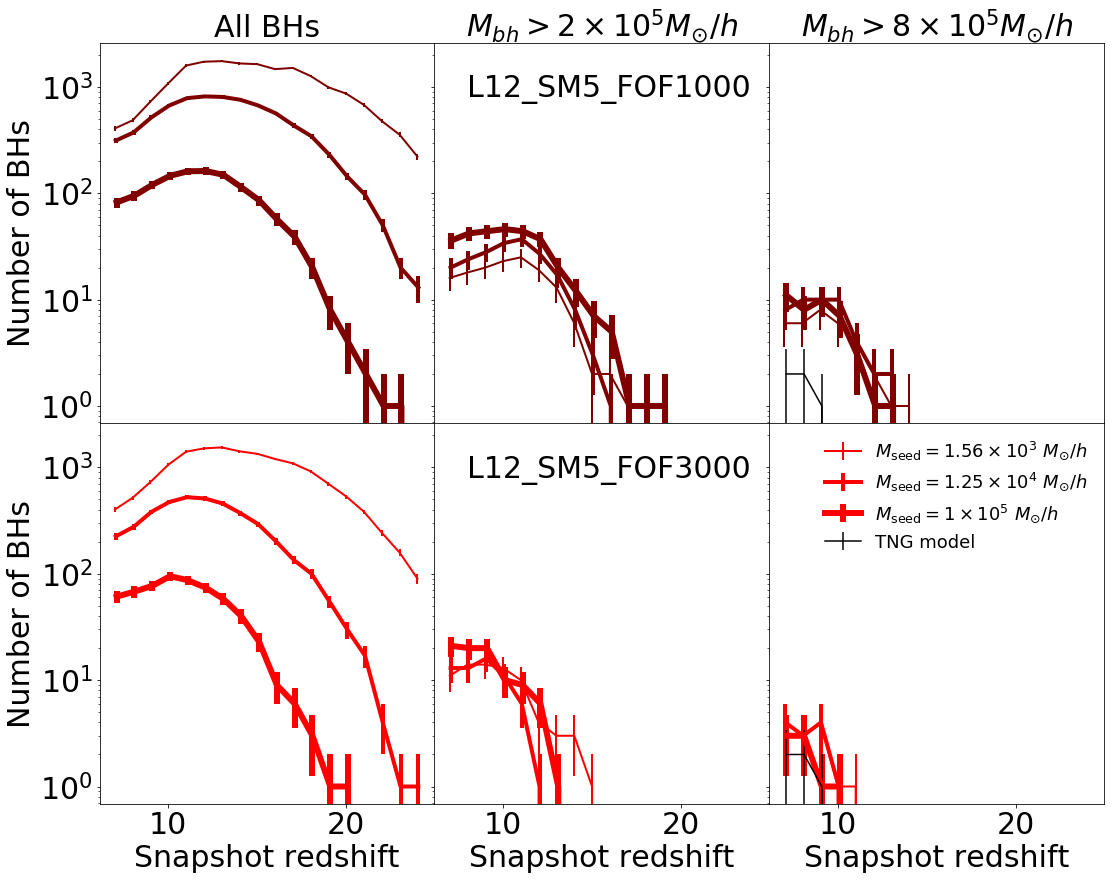} 
\caption{\textbf{Dependence of black hole count vs. redshift on $M_{\mathrm{seed}}$:} Number of black holes as a function of redshift for models with different $M_{\mathrm{seed}}$ at fixed $\tilde{M}_{\mathrm{sf,mp}}=5$ and $\tilde{M}_{h}$~(in region \texttt{ZOOM_REGION_z5}). Leftmost panel correspond to all BHs; middle and right panels correspond to black holes above fixed mass thresholds of $2\times10^5~M_{\odot}/h$ and $8\times10^5~M_{\odot}/h$ respectively. Upper and lower panels correspond to $\tilde{M}_{h}=10^3$ and $3\times10^3$ respectively. The black line in the 3rd panels corresponds to the TNG seed model applied to our zoom region. These are results for $L_{\mathrm{max}}=12$. Left to right panels correspond to increasing black hole mass thresholds. We find that the number of black holes is generally insensitive to $M_{\mathrm{seed}}$~(particularly at $z\sim7-10$) for mass thresholds substantially higher than the $M_{\mathrm{seed}}$ values.}
\label{No_of_BHs_seedmass_dependence_fig}

\begin{tabular}{cc}
\includegraphics[width=6cm]{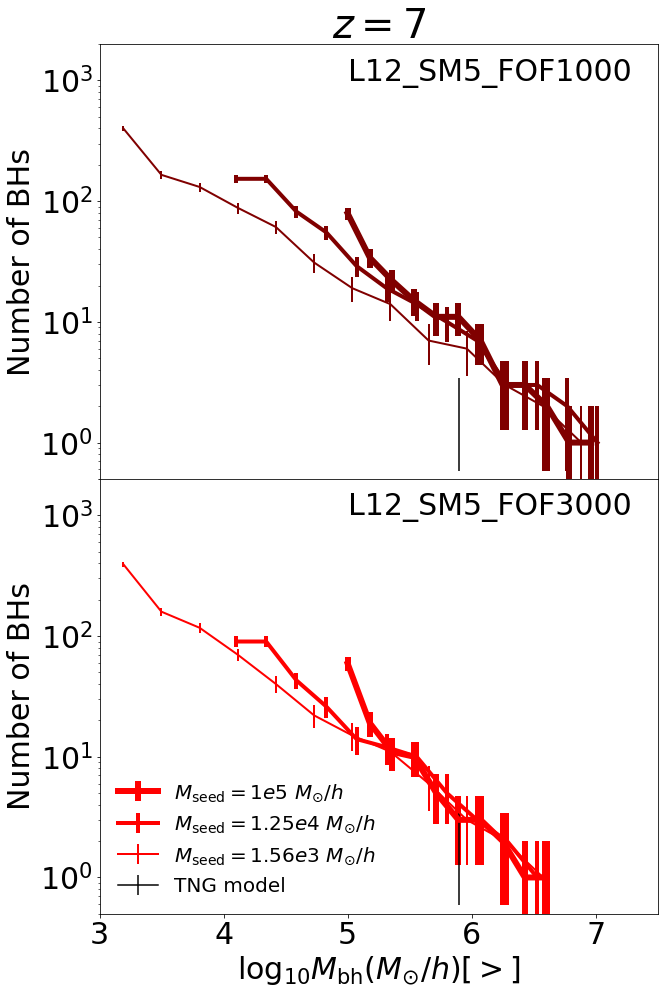} & \includegraphics[width=5.8cm]{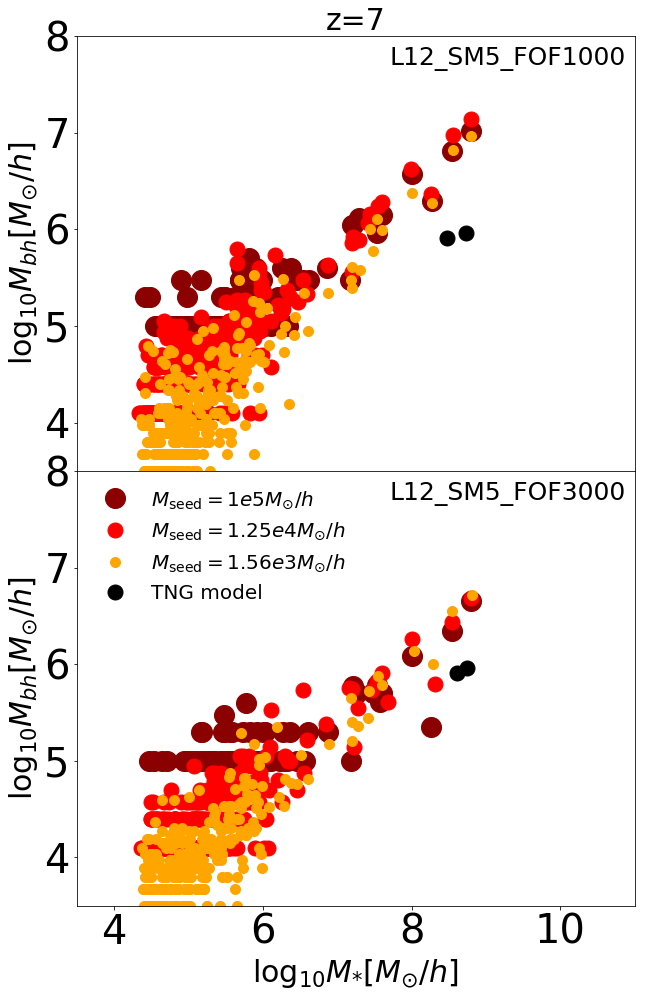}\\
\end{tabular}
\caption{\textbf{Dependence of cumulative mass functions and $M_{bh}-M_{*}$ relations on $M_{\mathrm{seed}}$:} Left and right panels show the cumulative mass functions and $M_{bh}-M_{*}$ relations for models with different $M_{\mathrm{seed}}$ at fixed $\tilde{M}_{\mathrm{sf,mp}}=5$ and $\tilde{M}_{h}$~(in region \texttt{ZOOM_REGION_z5}) at $z=7$. In the right panels, $M_*$ and $M_{bh}$ correspond to the total stellar mass and black hole mass of subhalos. Upper and lower panels correspond to $\tilde{M}_{h}=10^3$ and $3\times10^3$ respectively. Black color corresponds to the TNG model. We find that the massive end~($\sim3\times10^5-10^{7}~M_{\odot}/h$) of the black hole mass function and $M_{bh}-M_{*}$ relation is insensitive to $M_{\mathrm{seed}}$.}
\label{Dependence of cumulative mass functions on Mseed}
\end{figure*}

In this section, we investigate the masses of the SMBH populations produced 
by our seed models. Figure \ref{No_of_BHs_seedmass_dependence_fig} shows the number of black holes at each redshift snapshot. Note again that we do not present volume independent quantities such as comoving number densities because the overall zoom volume is not very well defined and is distorted by gravitational collapse. Also, recall from Section \ref{Target regions for high resolution zoom-ins} that we only include black holes within $1~\mathrm{Mpc}/h$ from the center of mass of the zoom volume, to avoid regions contaminated with low resolution DM particles. In general, we see that there is a steady increase in the number of black holes up to $z\sim11$, when black holes are rapidly forming and growing via mergers. At $z\sim7-11$ where seed formation begins to be suppressed by metal enrichment, merger events start to gradually decrease the number of black holes with time. 

We now focus on how these results are impacted by $M_{\mathrm{seed}}$. The leftmost panel of Figure \ref{No_of_BHs_seedmass_dependence_fig} shows the overall number of BHs; not surprisingly, there is a significantly higher number of black holes produced by models with lower mass seeds. In particular, $M_{\mathrm{seed}}=1.56\times10^3~M_{\odot}/h$ produces $\sim100$ times more black holes compared to $M_{\mathrm{seed}}=1\times10^5~M_{\odot}/h$ at $z\gtrsim15$; at $z\sim7-11$, the black hole counts for $M_{\mathrm{seed}}=1.56\times10^3~M_{\odot}/h$ is $7-10$ times higher than $M_{\mathrm{seed}}=1\times10^5~M_{\odot}/h$. We are also interested in the comparison of black hole counts above a minimum threshold mass for different $M_{\mathrm{seed}}$~(note that this minimum threshold mass is chosen to subsample the black hole populations, and is different from the intrinsic seed mass $M_{\mathrm{seed}}$ used during the simulation runs). The results of this  comparison depend crucially on the relative growth rates of low mass vs high mass seeds. Above a mass threshold of $2\times10^5~M_{\odot}/h$, $M_{\mathrm{seed}}=1\times10^5~M_{\odot}/h$ produces $\sim3$ times more black holes for $\tilde{M}_{\mathrm{h}}=10^3$, compared to $M_{\mathrm{seed}}=1.56\times10^3~M_{\odot}/h$. The impact seems to be somewhat weaker for $\tilde{M}_{\mathrm{h}}=3\times10^3$~(although that might also be partly due to fewer statistics). But overall, the impact of $M_{\mathrm{seed}}$ on black hole counts above a fixed mass threshold is relatively small. This is because lower mass seeds form earlier, which gives them enough time to grow~(via mergers) and catch up with the higher mass seeds that form later.

We also note in Figure \ref{No_of_BHs_seedmass_dependence_fig} that as the threshold mass increases to $8\times10^{5}~M_{\odot}/h$~(rightmost panels), the black hole counts for the different $M_{\mathrm{seed}}$ models become indistinguishable~(given the statistical uncertainties). Note also that these black holes are significantly larger than all the $M_{\mathrm{seed}}$ values that are being compared. To further explore the black hole abundances for even higher threshold masses, we show in Figure \ref{Dependence of cumulative mass functions on Mseed} the cumulative mass functions and $M_{bh}-M_*$ relations at $z=7$. We see that the black hole mass functions for the different models start to become similar for black holes $\sim3-4$ times heavier than the most massive seed; they continue to remain similar up to the most massive BHs~($\sim10^7~M_{\odot}/h$). We can even more clearly see this in the $M_{bh}-M_*$ relations~(Figure \ref{Dependence of cumulative mass functions on Mseed}: right panels), wherein the most massive black holes have almost the same masses for models with different $M_{\mathrm{seed}}$. This suggests that if seeds of different masses form in halos with commensurately different total mass and star forming, metal poor mass, they can grow into SMBHs with similar masses between  $\sim10^6-10^7~M_{\odot}/h$ by $z=7$. 
 
\begin{figure*}
\begin{tabular}{cc}
\includegraphics[width=6cm]{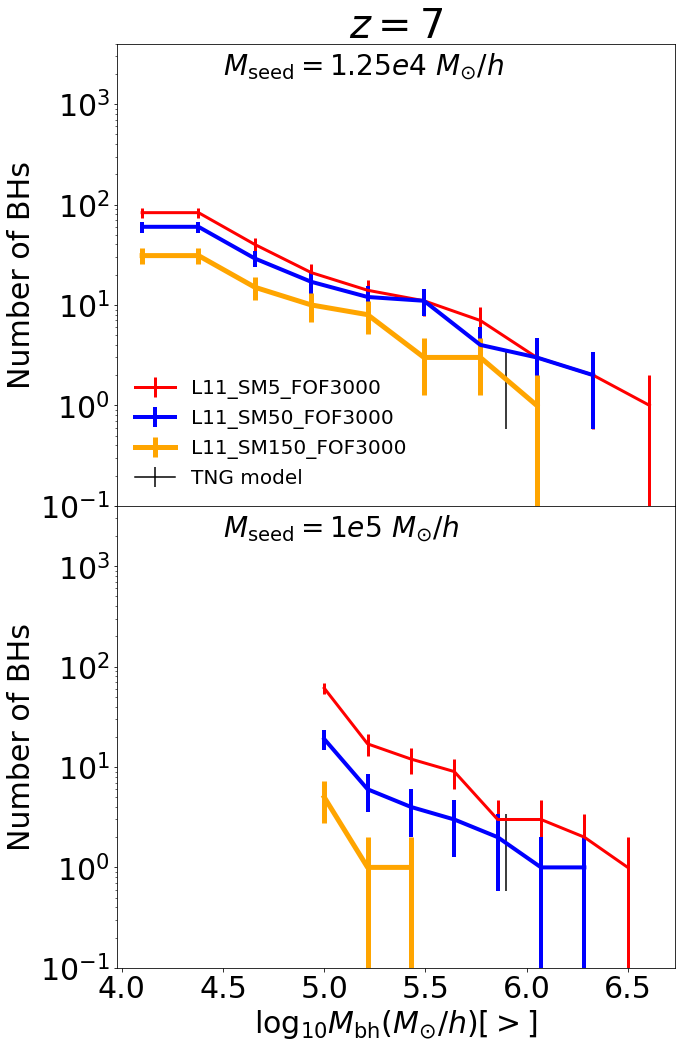} & \includegraphics[width=5.8cm]{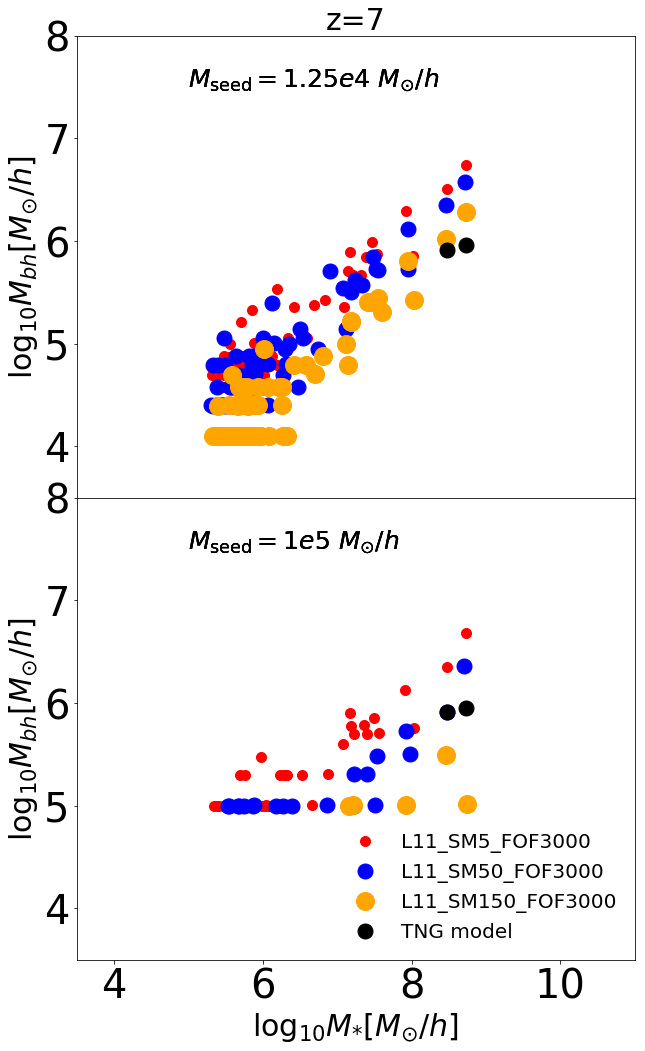}\\
\end{tabular}
\caption{\textbf{Dependence of cumulative mass functions and $M_{bh}-M_{*}$ relations on $\tilde{M}_{\mathrm{sf,mp}}$:} Left and right panels show the cumulative mass functions and $M_{bh}-M_{*}$ relations for models with different $\tilde{M}_{\mathrm{sf,mp}}$ at fixed $M_{\mathrm{seed}}$ and $\tilde{M}_{h}=3\times10^3$. In the right panels, $M_*$ and $M_{bh}$ correspond to the total stellar mass and black hole mass of subhalos. Upper and lower panels correspond to $M_{\mathrm{seed}}=1.25\times10^{4}~M_{\odot}/h$ and $1\times10^{5}~M_{\odot}/h$ respectively. The black color corresponds to predictions from the TNG model. For the most massive black holes at $z=7$, black hole masses are suppressed~(by a factor of $\sim8$ for $M_{\mathrm{seed}}=1.25\times10^{4}~M_{\odot}/h$) when $\tilde{M}_{\mathrm{sf,mp}}$ is increased~(from 5 to 150). The suppression is stronger for higher seed masses.}
\label{Dependence of cumulative mass functions on Msfmp}

\begin{tabular}{cc}
\includegraphics[width=6cm]{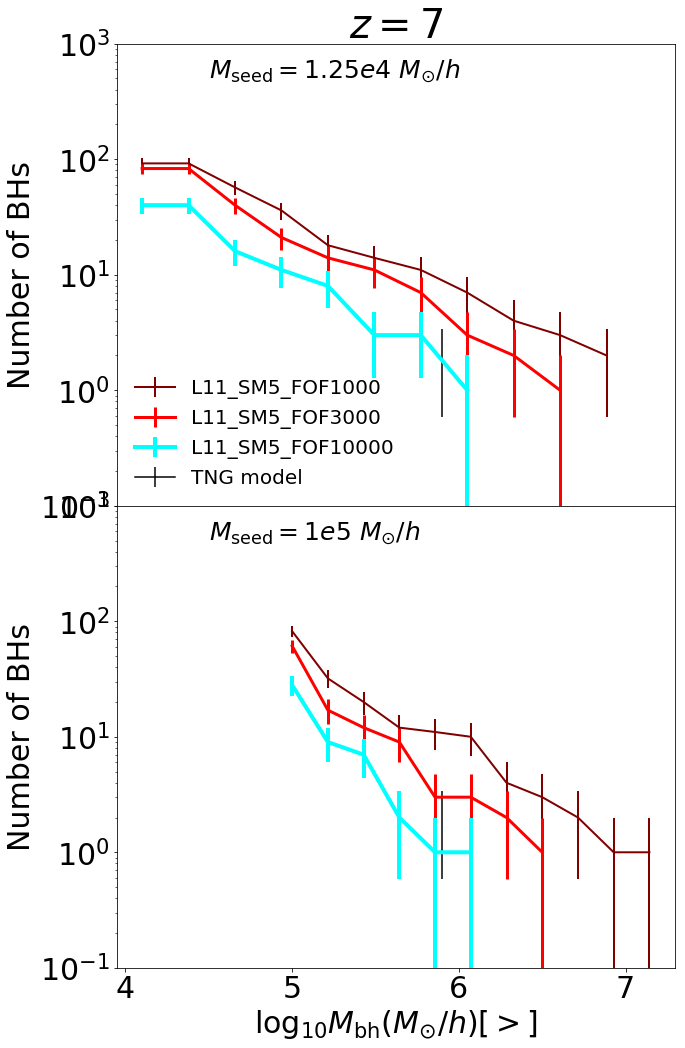} & \includegraphics[width=5.8cm]{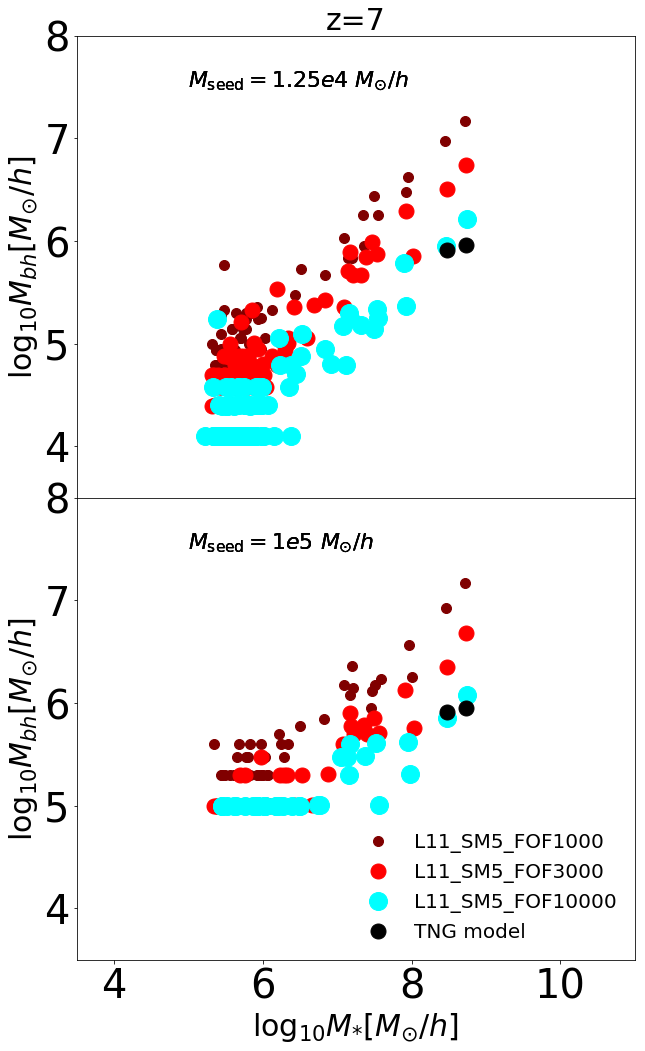}\\
\end{tabular}
\caption{\textbf{Dependence of cumulative mass functions and $M_{bh}-M_{*}$ relations on $\tilde{M}_{\mathrm{h}}$:} Left and right panels show the cumulative mass functions and $M_{bh}-M_{*}$ relations respectively for models with different $\tilde{M}_{\mathrm{h}}$ at fixed $M_{\mathrm{seed}}$ and $\tilde{M}_{\mathrm{sf,mp}}=5$. In the right panels, $M_*$ and $M_{bh}$ correspond to the total stellar mass and black hole mass of subhalos. Upper and lower panels correspond to $M_{\mathrm{seed}}=1.25\times10^{4}~M_{\odot}/h$ and $1\times10^{5}~M_{\odot}/h$ respectively. For the most massive black holes at $z=7$, black hole masses are suppressed~(by factors of $\sim10$) when $\tilde{M}_{\mathrm{h}}$ is increased~(from $10^3$ to $10^4$). The suppression is similar for both seed masses.}
\label{Dependence of cumulative mass functions on Mh}
\end{figure*}

$\tilde{M}_{\mathrm{sf,mp}}$ and $\tilde{M}_{h}$ have a more substantial impact~(compared to $M_{\mathrm{seed}}$) at the massive end~($\gtrsim10^6~M_{\odot}/h$) of the cumulative mass functions and the $M_*-M_{bh}$ relations~(shown in Figures \ref{Dependence of cumulative mass functions on Msfmp} and \ref{Dependence of cumulative mass functions on Mh}). At the most massive end, black hole masses are suppressed by factors of $\sim6$ when $\tilde{M}_{\mathrm{sf,mp}}$ is increased from $5-150$. When $\tilde{M}_{h}$ is increased by $10^3-10^4$, black hole masses are suppressed by factors of $\sim10$. Additionally, the impact $\tilde{M}_{\mathrm{sf,mp}}$ and $\tilde{M}_{h}$ on the mass functions are also qualitatively distinguishible. In particular, the impact of $\tilde{M}_{\mathrm{sf,mp}}$ on black hole abundances is largely uniform for the entire range of masses~($\sim10^4-10^7~M_{\odot}/h$) probed by these models. On the other hand, impact of $\tilde{M}_{h}$ is somewhat stronger at the massive end of the black hole mass function~(compared to the low mass end).

Lastly, we also compare our results with the TNG model, as applied to \texttt{ZOOM_REGION_z5}. These are shown as black lines / points throughout Figures \ref{No_of_BHs_seedmass_dependence_fig} to \ref{Dependence of cumulative mass functions on Mh}. Recall from Section \ref{Seed formation times} that when this model is applied to our zoom region, only two halos produce black hole seeds~($8\times10^5~M_{\odot}/h$) at $z\sim9$~(see Figures \ref{Seed_distributions_seed_dependence_fig}-\ref{Seed_distributions_FOF_dependence_fig}); they do not have any significant growth from $z\sim9$ to $z\sim7$. We now look at the black hole masses produced in these two halos using the gas-based seed models. These correspond to the most massive black holes in the right panels of Figures \ref{Dependence of cumulative mass functions on Mseed}, \ref{Dependence of cumulative mass functions on Msfmp} and \ref{Dependence of cumulative mass functions on Mh}. We see that our gas-based seed models tend to produce black hole masses $\sim2-10$ times higher than the TNG model.

\subsection{Merger rates}
\label{Merger rates}
\begin{figure*}
\includegraphics[width=13cm]{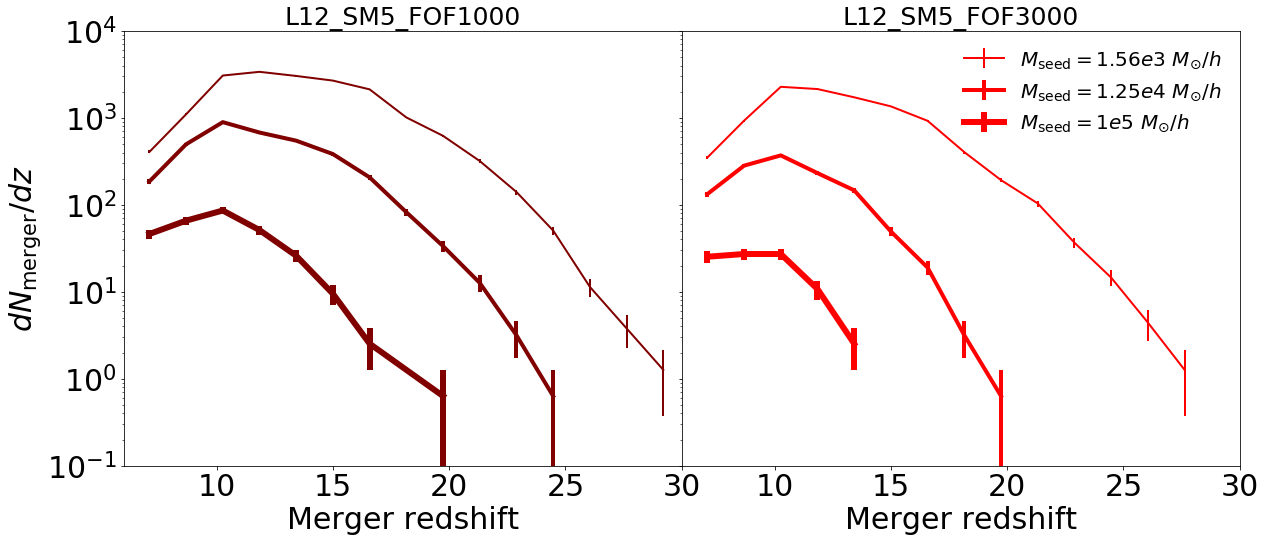} 
\caption{\textbf{Dependence of merger rates on $M_{\mathrm{seed}}$:} Merger rates of all black holes for models with different $M_{\mathrm{seed}}$ at fixed $\tilde{M}_{\mathrm{sf,mp}}=5$ and $\tilde{M}_{h}=3\times10^4$~(for \texttt{ZOOM_REGION_z5}). Left and right panels correspond to $\tilde{M}_{h}=10^3$ and $3\times10^3$ respectively. These are all run at $L_{\mathrm{max}}=12$. Not surprisingly, lower mass seeds produce higher numbers of mergers.  $M_{\mathrm{seed}}=1.56\times10^{3}~M_{\odot}/h$ produces $\sim10$ times more mergers at $z\sim7$ and $\sim1000$ times more mergers at $z\sim15$, compared to $M_{\mathrm{seed}}=1\times10^{5}~M_{\odot}/h$,.}
\label{All_Merger_rates_seedmass_dependence_fig}

\includegraphics[width=11cm]{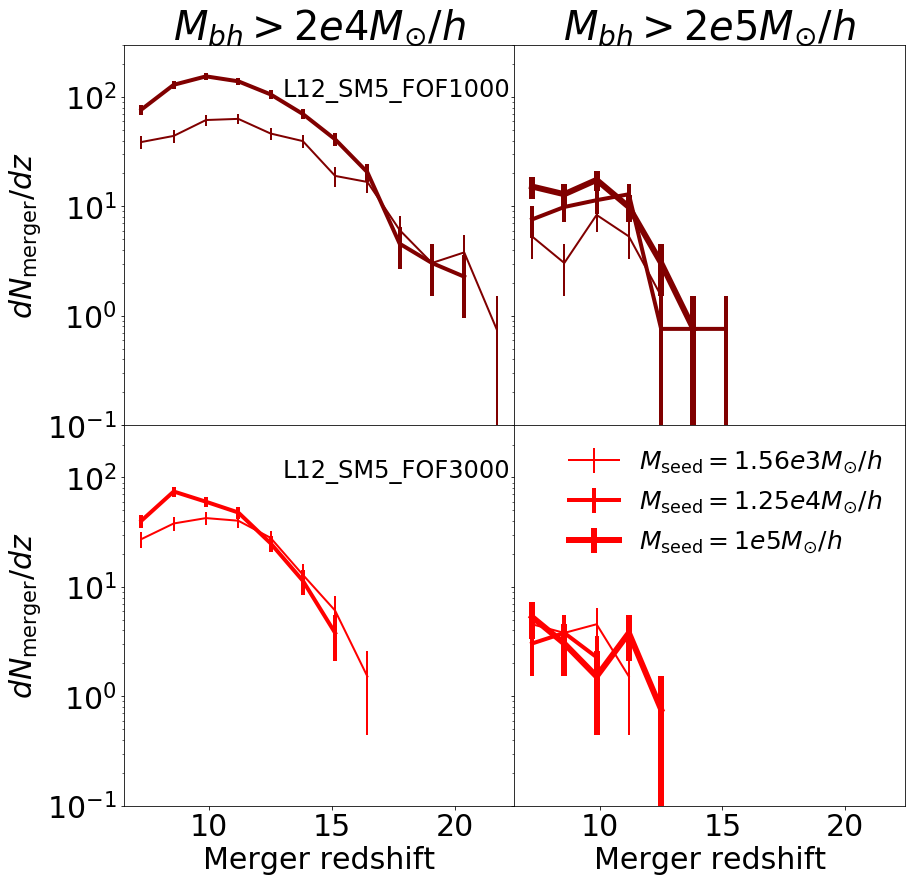} 
\caption{\textbf{Dependence of merger rates on $M_{\mathrm{seed}}$:} Merger rates of black holes above fixed mass thresholds for models with different $M_{\mathrm{seed}}$ at fixed $\tilde{M}_{\mathrm{sf,mp}}=5$ and $\tilde{M}_{h}$~(for \texttt{ZOOM_REGION_z5}). Upper and lower panels correspond to $\tilde{M}_{h}=10^3$ and $3\times10^3$ respectively. These are all run at $L_{\mathrm{max}}=12$. Here we show that when a black hole mass threshold is applied, the differences between the different $M_{\mathrm{seed}}$ values become substantially smaller~(up to factors of $\sim2-3$), compared to Figure \ref{All_Merger_rates_seedmass_dependence_fig}.}
\label{Merger_rates_seedmass_dependence_fig}
\end{figure*}

\begin{figure*}
\includegraphics[width=13cm]{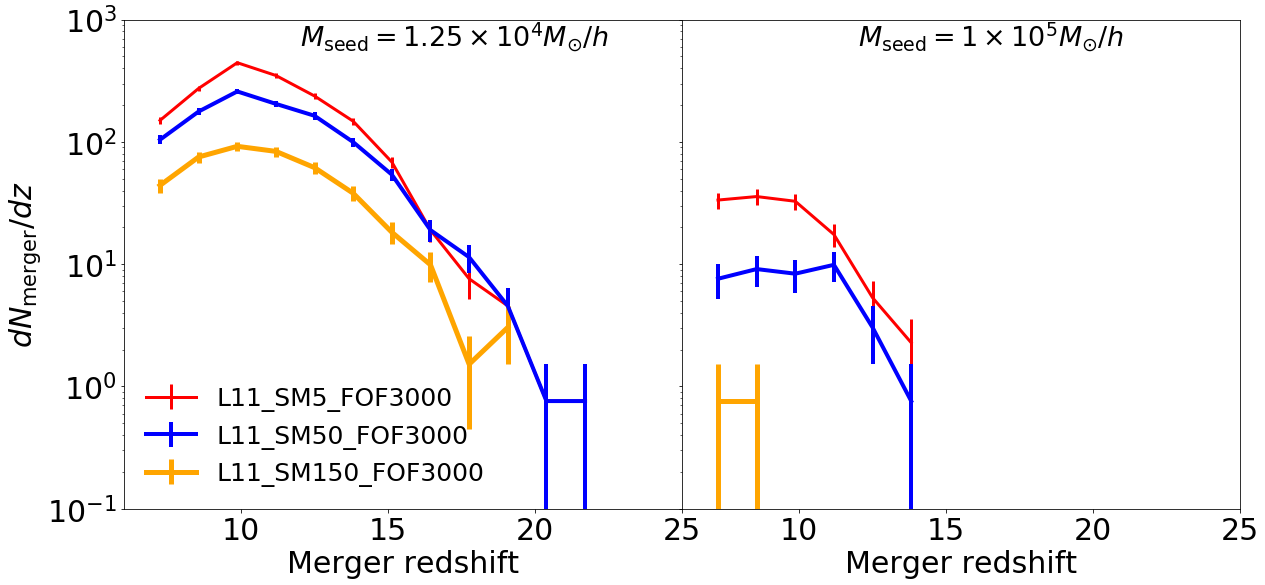} 
\caption{\textbf{Dependence of merger rates on  $\tilde{M}_{\mathrm{sf,mp}}$:} Merger rates of all black holes for models with different $\tilde{M}_{\mathrm{sf,mp}}$ at fixed $M_{\mathrm{seed}}$ and $\tilde{M}_{h}$~(for \texttt{ZOOM_REGION_z5}, $M_{\mathrm{seed}}=1.25\times10^{4}~M_{\odot}/h$). These are all run at $L_{\mathrm{max}}=11$. As we increase $\tilde{M}_{\mathrm{sf,mp}}$ from $5$ to $150$, the merger rates are suppressed by factors $\sim8$ for $M_{\mathrm{seed}}=1.25\times10^4~M_{\odot}/h$. The suppression is only slightly smaller at $z\gtrsim15$ compared to $z\sim7-11$. Lastly, the suppression is stronger for higher mass seeds.}
\label{Merger_rates_SM_dependence_fig}

\includegraphics[width=13cm]{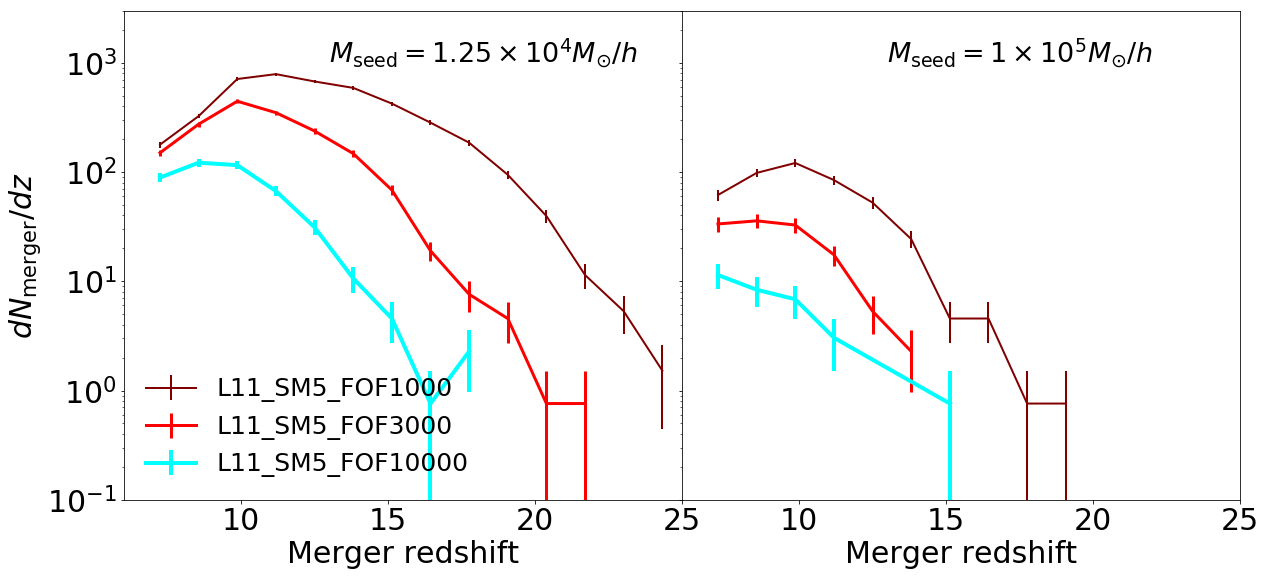} 
\caption{\textbf{Dependence of merger rates on $\tilde{M}_{h}$:} Merger rates of all black holes for models with different $\tilde{M}_{h}$ at fixed $\tilde{M}_{\mathrm{sf,mp}}$ and $M_{\mathrm{seed}}$~(for \texttt{ZOOM_REGION_z5}, $M_{\mathrm{seed}}=1.25\times10^{4}~M_{\odot}/h$). These are all run at $L_{\mathrm{max}}=11$. As we increase $\tilde{M}_{h}$ from $10^3$ to $10^4$, the merger rates have disproportionately larger suppression at higher redshifts. The suppression is by factors $\gtrsim100$ at $z\gtrsim15$, whereas at $z\sim7-11$, it is only by factors $\sim2-4$.}
\label{Merger_rates_FOF_dependence_fig}
\end{figure*}

In this section, we will quantify the impact of our seed model parameters on one of the most promising observable diagnostics of black hole seed models; i.e. black hole merger rates that could be probed with LISA. Before moving on, recall that two black holes are promptly merged when the distance between them is below the neighbor search radius of one of the black holes. Therefore, we are not able to probe the dynamics and finite timescale for black hole mergers 
below the resolution limit of the simulation. This would inevitably impact our merger rate predictions depending on the time delay between the simulated merger event and the ``actual merger event". For example, if we assume a fixed time delay for every merger, our predicted merger rates would be higher than the ``actual merger rate"~(i.e. merger rates inferred after accounting for the time delay) at epochs when the merger rate is increasing with time; conversely, the predicted merger rates would be lower than the ``actual merger rate" at epochs when the merger rate is decreasing with time. As discussed above, however, the complexity of zoom simulation volumes prevents us from making volume-independent merger rate predictions in any case. Thus, our primary focus here is the dependence of merger rates on different seed model parameters. 
The merger rates vs. redshift are shown in Figures \ref{All_Merger_rates_seedmass_dependence_fig}-\ref{Merger_rates_FOF_dependence_fig}. The merger rates tend to increase from $z\gtrsim15$ to $z\sim10$; they generally peak at around $z\sim10$. From $z\sim10$ to $z\sim7$, the merger rates start to gradually flatten and drop~(as the seeding events start also to drop). 

Figure \ref{All_Merger_rates_seedmass_dependence_fig} shows the impact of $M_{\mathrm{seed}}$ on the merger rates at fixed $\tilde{M}_{\mathrm{sf,mp}}$ and $\tilde{M}_{\mathrm{h}}$. As expected, models with lower mass seeds tend to produce more 
mergers, largely because lower mass seeds are overall more numerous. 
More specifically, $M_{\mathrm{seed}}=1.56\times10^{3}~M_{\odot}$ produces $\sim10$ times more mergers at $z\sim7$ and $\sim1000$ times more mergers at $z\sim15$, compared to $M_{\mathrm{seed}}=1\times10^{5}~M_{\odot}$. 

Next, we are interested in how the merger rates above a fixed black hole mass threshold vary with $M_{\mathrm{seed}}$; these are shown in Figure \ref{Merger_rates_seedmass_dependence_fig}. We find that the rate of mergers between black holes above a fixed mass varies much less with $M_{\mathrm{seed}}$ than does the global merger rate. 
This is not very surprising given that the overall black hole counts above fixed mass thresholds~(recall Figure \ref{No_of_BHs_seedmass_dependence_fig}) also show a similar amount of variation. We now take a closer look at different mass thresholds, starting with $>2\times10^4~M_{\odot}/h$~(Figure \ref{Merger_rates_seedmass_dependence_fig}: left panels) black holes. Here, we can compare results for $M_{\mathrm{seed}}=1.56\times10^3, 1.25\times10^4~M_{\odot}/h$. For $\tilde{M}_{\mathrm{h}}=1000$~(Figure \ref{Merger_rates_seedmass_dependence_fig}: upper panels), there is generally a higher number of mergers~(by factors up to $\sim3$) for $M_{\mathrm{seed}}=1.25\times10^{4}~M_{\odot}/h$, compared to $M_{\mathrm{seed}}=1.56\times10^{3}~M_{\odot}/h$. For mass thresholds of $2\times10^{5}~M_{\odot}/h$~(Figure \ref{Merger_rates_seedmass_dependence_fig}: right panels), we see similarly higher merger rates for $M_{\mathrm{seed}}=1\times10^{5}~M_{\odot}/h$ compared to $M_{\mathrm{seed}}=1.56\times10^3~M_{\odot}/h$~(albeit with fewer statistics). At higher $\tilde{M}_{\mathrm{h}}=3000$~(Figure \ref{Merger_rates_seedmass_dependence_fig}: lower panels), the variation with respect to $M_{\mathrm{seed}}$ seems to be slightly smaller.


Figure \ref{Merger_rates_SM_dependence_fig} shows 
how varying $\tilde{M}_{\mathrm{sf,mp}}$ for fixed $M_{\mathrm{seed}}=1.25\times10^4~M_{\odot}/h$ and 
$\tilde{M}_{\mathrm{h}}=3000$ impacts 
the overall merger rates. We can see that across the entire redshift range of $z\sim7-15$, the merger rates are suppressed by factors of $\sim8$ when $\tilde{M}_{\mathrm{sf,mp}}$ is increased from 5 to 150. At $z\gtrsim 15$, variations in $\tilde{M}_{\mathrm{sf,mp}}$ have a slightly smaller effect on the merger rates. 
In stark contrast, 
Figure \ref{Merger_rates_FOF_dependence_fig} 
demonstrates that the impact of $\tilde{M}_{\mathrm{h}}$ 
is stronger at higher redshifts. More specifically, at $z\sim7$ the merger rates are suppressed by factors of $\sim8$ when $\tilde{M}_{\mathrm{h}}$ is increased from $10^3$ to $10^4$; but the suppression is by factors $\gtrsim100$ for $z\gtrsim15$. 

Overall, we find that each of our model parameters have distinct imprints on the merger rates. First, $M_{\mathrm{seed}}$ naturally has the maximum impact on the merger rates amongst all the parameters, largely because it substantially increases the overall number of black holes. Next, the effects of $\tilde{M}_{\mathrm{h}}$ and $\tilde{M}_{\mathrm{sf,mp}}$ are dramatically different at different redshifts. The influence of $\tilde{M}_{\mathrm{sf,mp}}$ is relatively uniform whereas the impact of $\tilde{M}_{\mathrm{h}}$ is disproportionately stronger at higher redshifts. These distinctions are primarily governed by how our model parameters influence the seed formation times. These findings are useful in the context of LISA, which is expected to be able to detect $\gtrsim10^4~M_{\odot}/h$ black hole mergers all the way up to $z\sim15$. In particular, they shall be useful for interpreting any potential tension that can occur between future measurements of LISA merger rates and simulation predictions. For example, if the tension in the simulated vs observed merger rates is substantial at $z\gtrsim15$, this would hint that the typical halo masses for the seeds of the observed black holes are significantly different from the simulations. On the other hand, the simulated and observed merger rates may be  similar at $z\gtrsim15$ but diverge at lower redshifts~($z\sim7-11$); this would hint that the seeds of observed black holes may have formed in halos with similar total mass, but different star forming, metal poor mass, compared to simulations.

Finally, we also note that TNG model~(with $M_{\mathrm{seed}}=8\times10^5~M_{\odot}/h$) does not produce any merger events at $z\gtrsim7$ within our zoom volume. Our gas-based seed models also produce only $\sim1-4$ mergers of $>8\times10^5~M_{\odot}/h$ black holes at $z\gtrsim7$. Therefore the statistics are too limited to make any firm conclusions about the comparison of merger rate predictions for the TNG model vs gas-based seed prescriptions (above a fixed mass threshold of $8\times10^5~M_{\odot}/h$). 

\subsection{AGN luminosity functions}
\label{Luminosity Functions}
\begin{figure*}
\includegraphics[width=15cm]{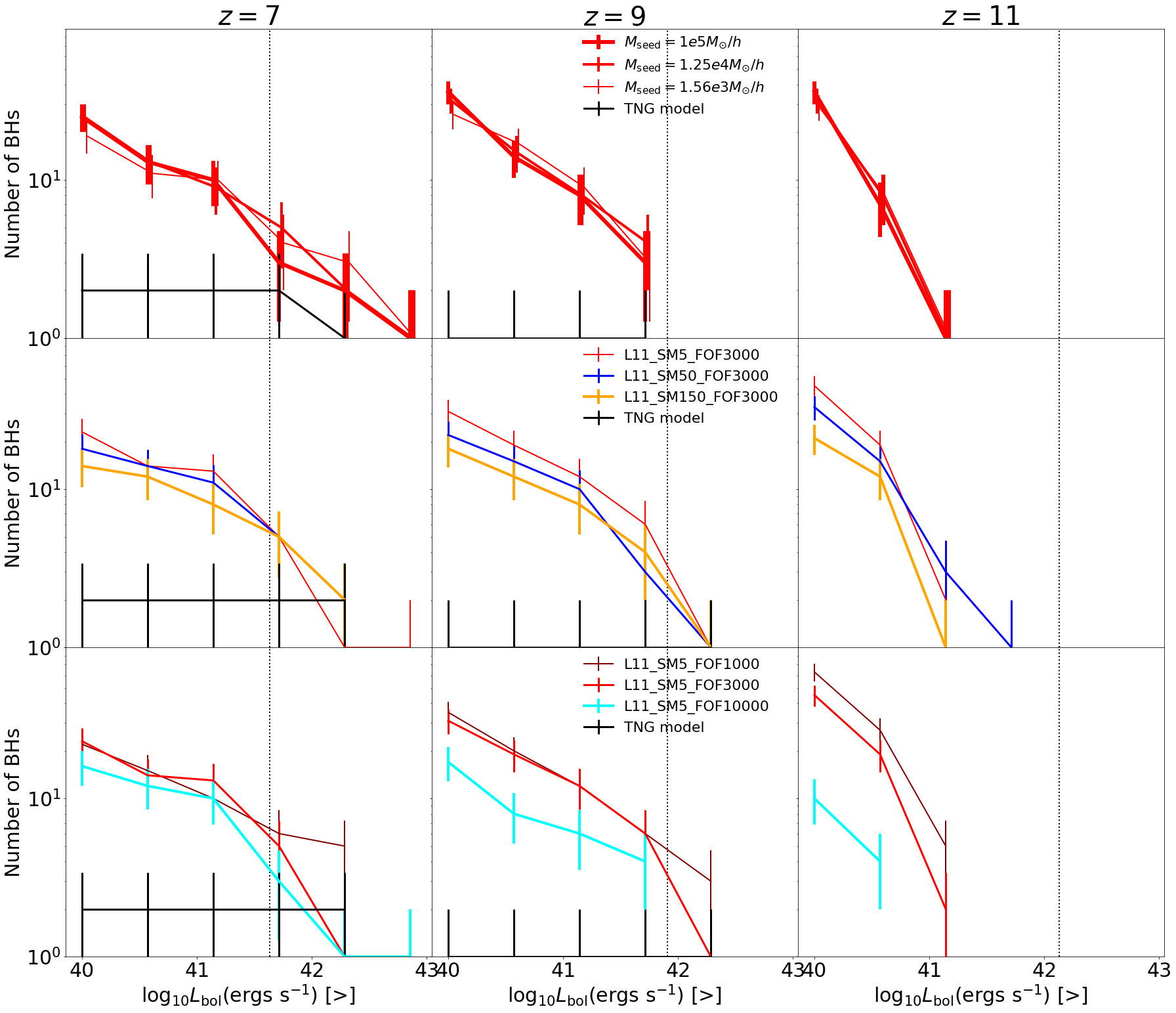} 
\caption{\textbf{Dependence of AGN luminosity functions on $M_{\mathrm{seed}}$, $\tilde{M}_{\mathrm{sf,mp}}$ and $\tilde{M}_{h}$:} Cumulative luminosity functions for different gas-based seed models as applied to \texttt{ZOOM_REGION_z5}. Upper panels correspond to models with different $M_{\mathrm{seed}}$ at fixed $\tilde{M}_{\mathrm{sf,mp}}=5$ and $\tilde{M}_{h}=3000$; these are run at $L_{\mathrm{max}}=12$. Middle panels show models with different $\tilde{M}_{\mathrm{sf,mp}}$ at fixed $M_{\mathrm{seed}}=1.25\times10^4~M_{\odot}/h$ and $\tilde{M}_{h}=3\times10^3$. Lower panels correspond to different $\tilde{M}_{h}$ at fixed $M_{\mathrm{seed}}=1.25\times10^4~M_{\odot}/h$ and $\tilde{M}_{sf,mp}=5$. The middle and lower panels are run at $\mathrm{L_{\mathrm{max}}}=11$. The black line corresponds to the TNG model. Vertical lines correspond to the detection limit of Lynx, assumed to be $1\times10^{-19}~\mathrm{ergs~cm^{-2}~s^{-1}}$ in the $2-10~\mathrm{keV}$ band for a survey area of $360~\mathrm{arcmin}^2$~\protect\citep{2020MNRAS.492.2535G}. The required bolometric correction is adopted from \protect\cite{2007MNRAS.381.1235V}. The variation in the luminosity functions is only up to factors of $\sim2-3$; therefore, it may be challenging for Lynx to be able to distinguish between these models.}
\label{Luminosity_functions_all_dependence_fig}
\end{figure*}

In this section, we focus our attention on the AGN luminosity functions shown in Figure \ref{Luminosity_functions_all_dependence_fig} for \texttt{ZOOM_REGION_z5}. Here too~(as with the black hole masses), black holes are confined to within $1~\mathrm{Mpc}/h$ from the center of mass of the zoom volume to avoid regions contaminated with low resolution DM particles. We compute the bolometric luminosities of the black holes from their mass accretion rates using Eq.~(\ref{bol_lum_eqn}). Our gas-based seed models produce AGN up to bolometric luminosities as high as $\sim10^{44}~\mathrm{ergs/s}$. In comparison, the TNG model has two black holes at $z=7$ with luminosities of $\sim10^{42}~\mathrm{ergs/s}$. However, these AGN are still 2-3 orders of magnitude fainter than the observed population of $z>6$ quasars. This is not unexpected given that the largest black holes that we form in \texttt{ZOOM_REGION_z5} have masses ranging from $\sim10^6-10^7~M_{\odot}/h$~(depending on the seed model), which is $\sim10-100$ times smaller than the typical masses of the first quasars. More generally, \texttt{ZOOM_REGION_z5} does not probe the observed high-z quasar regime because it is selected from relatively small parent volume of ~$(25~\mathrm{Mpc}/h)^3$. In comparison, simulations that do successfully produce a few first quasars are $\sim4000$ times larger in volume~\citep{2018MNRAS.474..597T,2018MNRAS.481.4877N,2020MNRAS.499.3819M}. Therefore, probing the observed high-z quasar regime is beyond the scope of this paper, and we reserve this aspect for future work. The following paragraphs focus on the effect of our seed parameters on the luminosity functions.

We quantify the influence of our gas-based seed models at the detection limit of Lynx, which is assumed to be $1\times10^{-19}~\mathrm{ergs~cm^{-2}~s^{-1}}$ in the $2-10~\mathrm{keV}$ band for a survey area of $360~\mathrm{arcmin}^2$~\citep{2020MNRAS.492.2535G}. To convert this into a bolometric luminosity limit, we assume the bolometric correction derived in \cite{2007MNRAS.381.1235V}. We focus on Lynx because it is one of the only planned missions that is deep enough to potentially probe our simulated luminosities. The impact of $M_{\mathrm{seed}}$ on the luminosity functions is overall small compared to statistical uncertainties~(Poisson errors) in our predictions~(see top panels of Figure \ref{Luminosity_functions_all_dependence_fig}). This is infact true all the way down to $10^{40}~\mathrm{ergs/s}$. Therefore, the differences in black hole abundances for models with different $M_{\mathrm{seed}}$~(seen in Figure \ref{No_of_BHs_seedmass_dependence_fig}) are largely contributed by very low luminosity~($\lesssim10^{40}~\mathrm{ergs/s}$) black holes. The luminosity functions do vary somewhat with $\tilde{M}_{\mathrm{sf,mp}}$ and $\tilde{M}_{\mathrm{h}}$~(see middle and bottom panels of Figures \ref{Luminosity_functions_all_dependence_fig}). For $M_{\mathrm{seed}}=1.25\times10^{4}~M_{\odot}/h$, the luminosity functions at the Lynx limit are suppressed by factors of $\sim2-3$ from $\tilde{M}_{\mathrm{sf,mp}}=5$ to $150$. Varying $\tilde{M}_{h}$ from $10^3$ to $10^4$ also suppresses the luminosity functions by factors of $\sim2-3$ at $z=7,9$. For higher seed masses~(not shown in Figure \ref{Luminosity_functions_all_dependence_fig}), we report broadly similar trends but the statistical uncertainties are too high to make robust conclusions. Overall, while these variations are noteworthy, they are still relatively small~(compared to the influence on merger rates, for instance). These variations are also somewhat smaller than the typical uncertainties in current measurements of high-z luminosity functions~($\sim10-100~\mathrm{dex}$ at the faint end, see Figure 5 of \citealt{2020MNRAS.495.3252S}). Furthermore, the luminosity functions may also be influenced by other aspects of our model such as black hole accretion and feedback. Therefore, we suspect that it will be challenging for luminosity function constraints from future surveys such as JWST and Lynx to be able to distinguish between our seed models. We can put these results in context of prior work done using semi-analytic modeling. It is noteworthy that our conclusions somewhat differ from \cite{2018MNRAS.481.3278R} who find that Lynx may be able to put constraints on seed models; however, this could be partly because their seeding and accretion prescriptions are widely different from ours. At the same time, \cite{2020MNRAS.492.2535G} finds that the $z\sim7-12$ luminosity functions are not sensitive to their choice of seed masses ranging from $10-10^5~M_{\odot}/h$, consistent with our findings. In any case, recent work by \cite{2017ApJ...838..117N} hints that multi wavelength spectral observations from high-z AGN and their host galaxies~(using JWST) may have substantially higher constraining power compared to the overall bolometric luminosities~(we plan to investigate this using our simulations in a future study). This emphasizes the importance of continued work on finding electromagnetic signatures of black hole seeds.

\begin{figure}
\includegraphics[width=8cm]{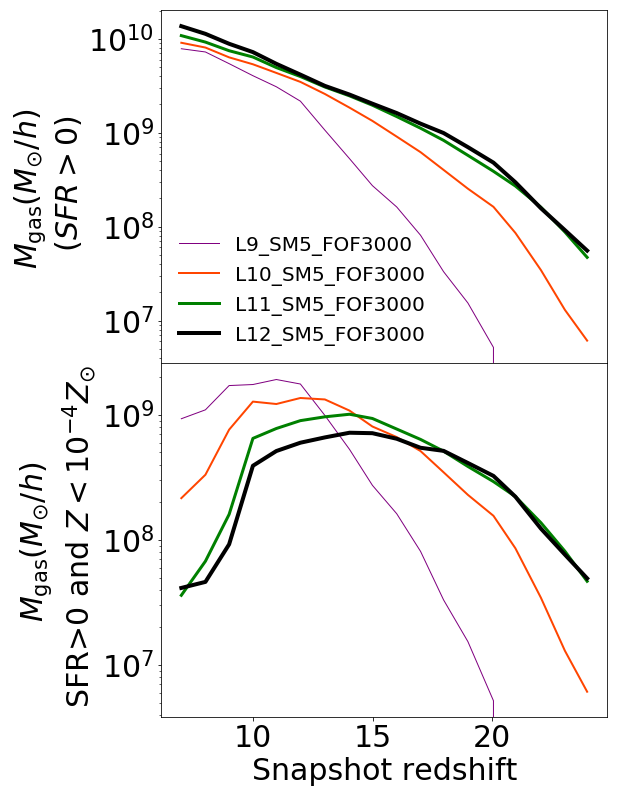} 
\caption{\textbf{Top Panel:} The evolution of the total mass of star forming gas over the entire \texttt{ZOOM_REGION_z5}~(for the same simulations as 
in Figures \ref{density_2dplot_fig} and \ref{metallicity_2dplot_fig}). The star formation evolution approaches convergence at $L_{\mathrm{max}}=11,12$. \textbf{Bottom Panel:} The evolution of the total mass of star forming, metal poor~($Z<10^{-4}~Z_{\odot}$) gas over the entire \texttt{ZOOM_REGION_z5}. Following the star formation, the metal enrichment evolution also approaches convergence at $L_{\mathrm{max}}=11,12$. However, the convergence is slower at $z\lesssim17$, wherein the $L_{\mathrm{max}}=12$ still has a notably faster metal enrichment compared to $L_{\mathrm{max}}=11$. This can also be visually seen in Figure \ref{metallicity_2dplot_fig}.}
\label{StarFormationMetalEnrichmentevolution_fig}
\end{figure}

\section{Resolution convergence}
\label{Resolution convergence}

\begin{figure*}

\includegraphics[width=15cm]{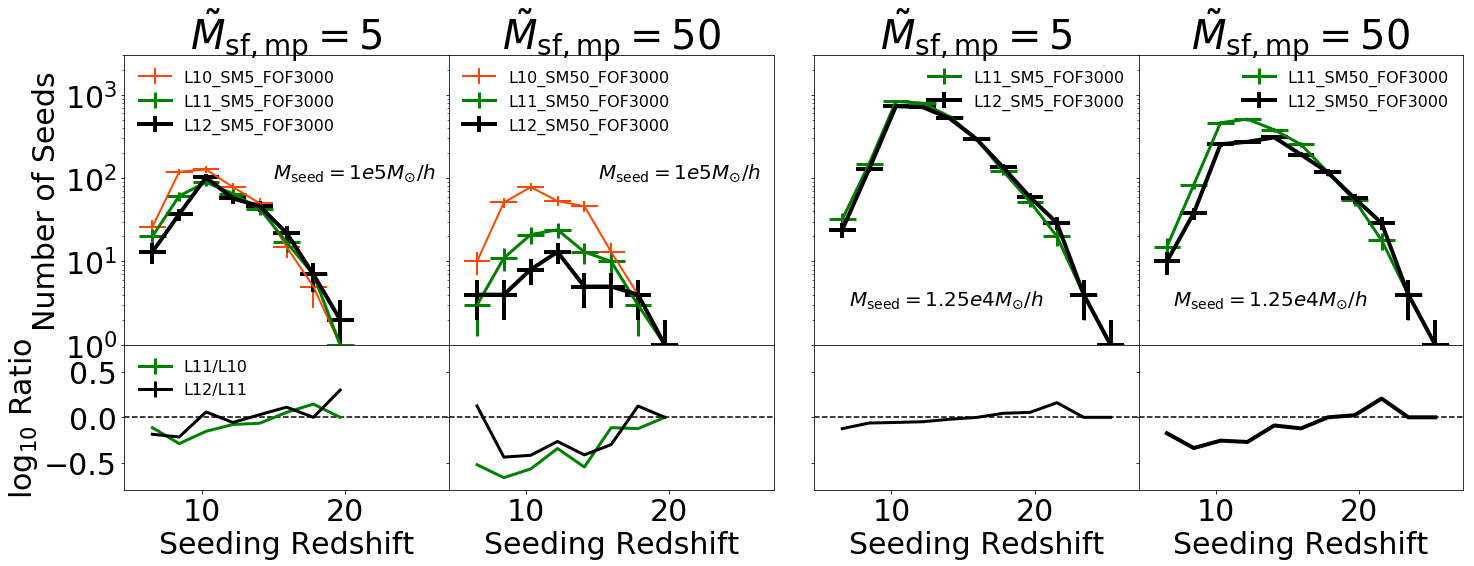} 

\caption{\textbf{Resolution convergence tests:} Distribution of seeding times for simulation boxes of different resolutions~(different colors) at fixed set of seed parameters~($M_{\mathrm{seed}}$, $\tilde{M}_{\mathrm{sf,mp}}$, $\tilde{M}_{h}$) for \texttt{ZOOM_REGION_z5}. Left and right panels correspond to $\tilde{M}_{\mathrm{sf,mp}}=5$ and 50 respectively. Upper and lower panels show seed masses of $1.25\times10^{4}~M_{\odot}/h$ and $1.56\times10^{3}~M_{\odot}/h$  respectively.  $\tilde{M}_{h}=3\times10^3$ in all the panels.  The distributions approach convergence for $L_{\mathrm{max}}\geq 11$. 
But note also that at $z<17$, the rate of convergence is slightly lower for higher values of $\tilde{M}_{\mathrm{sf,mp}}$ and $M_{\mathrm{seed}}$.}
\label{seed_times_resolution_convergence_fig}
\includegraphics[width=15cm]{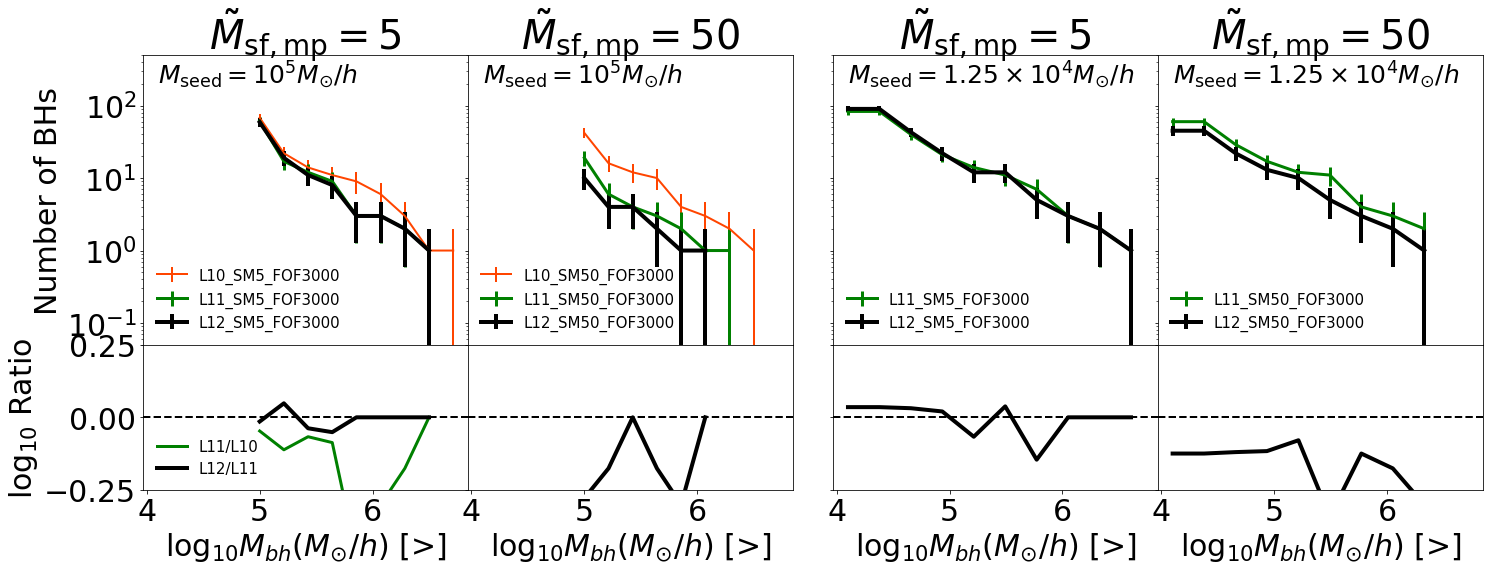}

\caption{\textbf{Resolution convergence tests:} Cumulative mass functions at $z=7$  for simulation boxes of different resolutions~(different colors) at fixed set of seed parameters~($M_{\mathrm{seed}}$, $\tilde{M}_{\mathrm{sf,mp}}$, $\tilde{M}_{h}$) for \texttt{ZOOM_REGION_z5}. $M^{\mathrm{group}}_*$ and $M^{\mathrm{group}}_{bh}$ are stellar masses and black hole masses of halos. $M_{bh}(>)$ is the minimum black hole mass above which the black hole counts are plotted in the right panels. The model parameters are the same as that of Figure \ref{seed_times_resolution_convergence_fig}~(see legends). $M_*-M_{bh}$ relations and cumulative mass functions approach convergence with increasing resolution. The rate of convergence is slightly lower for higher values of $\tilde{M}_{\mathrm{sf,mp}}$ and $M_{\mathrm{seed}}$.}
\label{SMBHM_mass_function_resolution_tests}

\includegraphics[width=15cm]{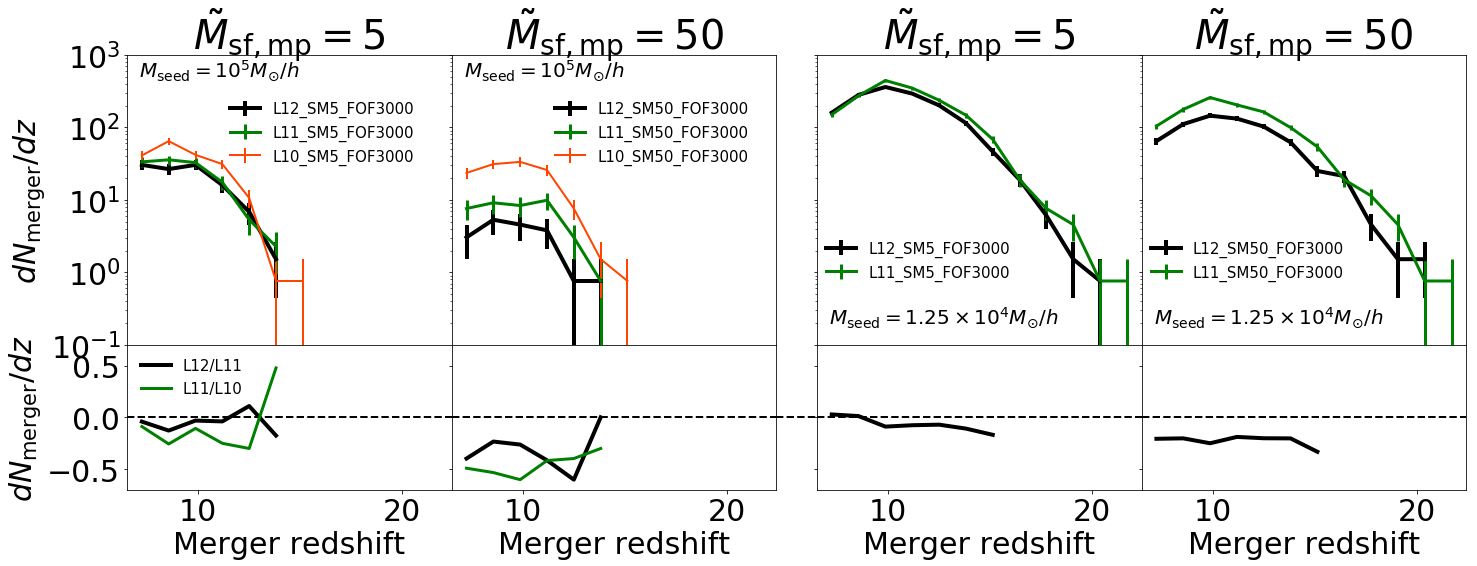}

\caption{\textbf{Resolution convergence tests:}: black hole merger rates for simulation boxes of different resolutions~(different colors) at fixed set of seed parameters~($M_{\mathrm{seed}}$, $\tilde{M}_{\mathrm{sf,mp}}$, $\tilde{M}_{h}$) for \texttt{ZOOM_REGION_z5}. $L_{\mathrm{bol}}(>)$ is the threshold bolometric luminosity. The model parameters are the same as that of Figure \ref{seed_times_resolution_convergence_fig}~(see legends). The merger rates approach convergence with increasing resolution. The rate of convergence is slightly lower for higher values of $\tilde{M}_{\mathrm{sf,mp}}$ and $M_{\mathrm{seed}}$.}
\label{luminosity_function_merger_rates_resolution_tests}

\end{figure*}

In this section, we examine the resolution convergence of the results presented in this work. Recall that the black hole seeding essentially depends on the amount of star forming, metal poor gas mass in halos as well as total mass of the halos. The total halo mass is known to converge well even at significantly lower resolutions compared to our zoom regions~\citep{2007ApJ...671.1160L}. Therefore, the resolution convergence of our final black hole populations is primarily driven by the resolution convergence of star formation and metal enrichment. We first revisit Figures \ref{density_2dplot_fig} and \ref{metallicity_2dplot_fig} which show the density and metallicity evolution at different resolutions i.e. $L_{\mathrm{max}}=9,10,11,12$. We see that with increasing $L_{\mathrm{max}}$, we are able to resolve higher densities on smaller scales. This leads to earlier onset of star formation and metal enrichment at higher resolutions. This is easily seen in Figure \ref{metallicity_2dplot_fig} wherein for $L_{\mathrm{max}}=11~\&~12$, the first metal enriched regions appear before $z=20$. In contrast, for $L_{\mathrm{max}}=9~\&~10$, the first metal enriched regions do not appear until after $z=20$. However, as we shall quantitatively demonstrate in the remaining paragraphs, our black hole seed models are reasonably well converged by $L_{\mathrm{max}}\geq11$. 

In Figure \ref{StarFormationMetalEnrichmentevolution_fig}, we quantify the resolution convergence of star formation and metal enrichment by showing the evolution of the total star forming gas mass~(top panel) and star forming, metal poor gas mass~
(bottom panel) of all gas cells in the zoom region at $L_{\mathrm{max}}=9,10,11,12$~(recall that metal poor gas cells are defined to have metallicity $Z<10^{-4}~Z_{\odot}$). We can see that the increase in star forming gas mass happens earlier at higher resolutions between $L_{\mathrm{max}}=9-11$, but star formation is well converged at $L_{\mathrm{max}}=11$ \& 12. 

The resolution convergence of  metal enrichment can be seen in Figure \ref{StarFormationMetalEnrichmentevolution_fig}~(bottom panel), where the star forming, metal poor gas mass starts to drop at $z\lesssim12-15$. At higher $L_{\mathrm{max}}$, the metal enrichment begins sooner, and the resulting drop in star forming, metal poor gas mass starts to occur at earlier redshifts. In particular, the drop occurs at $z\sim 15$ for $L_{\mathrm{max}}=12$ and at $z\sim 12$ for $L_{\mathrm{max}}=9$. Additionally, the drop also tends to become steeper at higher $L_{\mathrm{max}}$; this is an indicator of faster metal enrichment at higher $L_{\mathrm{max}}$.  At $z\sim7$, the star forming, metal poor gas mass is $\sim30$ times lower for $L_{\mathrm{max}}=11$ compared to $L_{\mathrm{max}}=9$. However, going from $L_{\mathrm{max}}=11$ to $L_{\mathrm{max}}=12$, the decrease is only by a factor $\sim1.5$. This is clear evidence that the metal enrichment exhibits resolution convergence. However, we do note the star forming, metal poor gas mass converges more slowly than the star forming gas mass. This implies that the convergence in metal enrichment is slower than that of star formation. We shall look at how this affects the convergence properties of black hole populations in the following paragraphs.  

Figure \ref{seed_times_resolution_convergence_fig} shows the distributions of seeding times~(redshifts) at different resolutions. 
We are able to perform resolution convergence tests across $L_{\mathrm{max}}=10-12$ for $M_{\mathrm{seed}}=1\times10^{5}~M_{\odot}/h$, and across $L_{\mathrm{max}}=11-12$ for $M_{\mathrm{seed}}=1.25\times10^{4}~M_{\odot}$.
Let us start with $\tilde{M}_{\mathrm{sf,mp}}=5$. At $M_{\mathrm{seed}}=1\times10^{5}~M_{\odot}$, we find that differences in the distributions between successive values of $L_{\mathrm{max}}$, becomes smaller at higher resolutions~(see lower panel of 1st column); between $L_{\mathrm{max}}=11,12$, the agreement is to within $\sim25\%$ at $z\gtrsim10$. At $M_{\mathrm{seed}}=1.25\times10^{4}~M_{\odot}/h$, we find even better agreement~($\lesssim 10\%$) between $L_{\mathrm{max}}=11,12$. These are overall strong signatures of resolution convergence. At $z\sim7-11$ however, the convergence is weaker 
and the number of seeds 
differs by factors $\sim2$ between $L_{\mathrm{max}}=12$ vs. $11$. Notice also that at $z\lesssim11$, the formation of new seeds begins to slow down as halos become increasingly metal-enriched. 
This strongly suggests that the weaker 
resolution convergence at $z\sim7-11$ is driven by the onset of metal enrichment~(and the fact the metal enrichment converges more slowly 
than star formation). 

If we increase $\tilde{M}_{\mathrm{sf,mp}}$ to 50~(see again Figure \ref{seed_times_resolution_convergence_fig}: 2nd and 4th columns panels), the differences between $L_{\mathrm{max}}=10,11,12$ start to appear even earlier~($z\lesssim15$), and the level of convergence is poorer. This reflects the earlier onset of metal enrichment in the higher-resolution runs, which places more stringent constraints on seed formation sites when a larger reservoir of star-forming, metal-poor gas is required. Overall, despite the varying rates of convergence for different parameter values, the important finding here is that our rates of seed formation indeed exhibit resolution convergence. Since $\tilde{M}_{\mathrm{sf,mp}}=5$ produces the fastest resolution convergence, we used this as our ``baseline" model and scanned for model variants around it in Section \ref{Impact of seed parameters on the BH population}. Moreover, note that in this work, we are more interested in variations of the various quantities~(BH luminosity functions, merger rates etc.) with respect to changing parameters, instead of the actual values for these quantities. These variations are not significantly affected by the decrease in convergence rate. 

We now look at how 
the above trends impact the convergence of the observable statistics of black hole populations. Figures \ref{SMBHM_mass_function_resolution_tests} and \ref{luminosity_function_merger_rates_resolution_tests} show the mass functions~(at $z=7$) and merger rates respectively. Let us start with $M_{\mathrm{seed}}=1\times10^{5}~M_{\odot}/h$. Here, we find that our predictions are indeed converging; i.e., differences between $L_{\mathrm{max}}=11~\mathrm{vs}~12$ predictions are smaller than that of $L_{\mathrm{max}}=10~\mathrm{vs}~11$; this is expected based on results from the previous paragraph. The mass functions converge to within factors of $\sim1.3$ and $\sim2$ for $\tilde{M}_{\mathrm{sf,mp}}=5$ and $50$ respectively by $L_{\mathrm{max}}=11$. The merger rates converge to within factors of $\sim2-4$ for $\tilde{M}_{\mathrm{sf,mp}}=5$ and $50$ by $L_{\mathrm{max}}=11$~(albeit 
with very limited statistics for the latter). Resolution convergence is even better for 
$M_{\mathrm{seed}}=1.25\times10^{4}~M_{\odot}/h$; this is consistent with expectations based on results from the previous paragraph and Figures \ref{StarFormationMetalEnrichmentevolution_fig} and \ref{seed_times_resolution_convergence_fig}.

Our analysis of resolution convergence reveals the following key findings:
\begin{itemize}
\item Our seed models produce black hole statistics~(mass functions, luminosity functions and merger rates) that are reasonably well converged by $L_{\mathrm{max}}=11$; this is largely driven by the resolution convergence of the star formation and metal enrichment.

\item Our results converge more slowly with resolution for larger values of $\tilde{M}_{\mathrm{sf,mp}}$ or $M_{\mathrm{seed}}$. 
This is due to a combination of two effects: 1) the resolution convergence of metal enrichment is slower than that of star formation, and 2) increasing $\tilde{M}_{\mathrm{sf,mp}}$ or 
$M_{\mathrm{seed}}$ pushes the seed formation to occur in regions with earlier metal enrichment.   
\end{itemize}

\section{Summary and Conclusions}
\label{Summary and Conclusions}
In this work, we perform a systematic study to characterize the outcome of different black hole seed models using cosmological hydrodynamic zoom simulations. To this end, we built black hole models wherein the seed formation explicitly occurs 
in halos with star forming, metal poor gas and sufficiently deep potentials. All the other aspects of the galaxy formation model~(which include star formation, feedback, metal enrichment) are identical to that of \texttt{IllustrisTNG}. Our main goal was to quantify the influence of various seed parameters on key SMBH observables such as the black hole masses,  luminosities and merger rates. The key parameters of our seed models were namely: \begin{enumerate} \item $M_{\mathrm{seed}}$: Black hole seed mass. \item $\tilde{M}_{\mathrm{sf,mp}}$: Minimum threshold of star forming, metal poor~($Z<10^{-4}~Z_{\odot}$) gas mass~(in units of $M_{\mathrm{seed}}$) in a halo. Only halos above this threshold are seeded with black holes.  \item $\tilde{M}_{h}$: Minimum threshold of total halo mass~(in units of $M_{\mathrm{seed}}$). Only halos above this threshold are seeded with black holes. \end{enumerate} 

We demonstrate that our model predictions approach resolution convergence in the zoom region; the convergence rate depends somewhat on the model parameters, as discussed in Section \ref{Resolution convergence}, but our results are reasonably converged for zoom resolutions of $L_{\mathrm{max}} \geq 11$. Thus, we use this resolution for most of our analysis. 

In our gas-based seed models, the onset of seed formation, driven by halo growth and the onset of star formation, occurs 
at redshifts ranging from $\sim12-25$~(depending on the model parameters). 
The rate of seed formation increases until its peak at $z\sim11$, after which 
metal enrichment starts to suppress the seeding. 
Different seed parameters have a qualitatively distinct influence on seed formation. Owing to the scaling of the gas and halo mass thresholds with $M_{\mathrm seed}$ in our models, higher $M_{\mathrm{seed}}$ values cause 
the more massive seeds to form in halos with higher star forming, metal free mass and total halo mass. Increasing $M_{\mathrm{seed}}$ from $1.56\times10^3-1\times10^6~M_{\odot}/h$ delays the onset of seeding and suppresses the number of seeds formed in each snapshot by up to factors of $\sim10$ and $\sim100$ for $z\sim7-11$ and $z\gtrsim11$, respectively. 
Increasing $\tilde{M}_{\mathrm{sf,mp}}$ from $5-150$ has very little effect on seed formation at $z\gtrsim15$, but higher $\tilde{M}_{\mathrm{sf,mp}}$ values suppress seed formation by factors up to $\sim 8$ at $z\lesssim15$, 
Higher $\tilde{M}_{\mathrm{h}}$ causes seeds of a given mass to form in halos with higher total mass~(or deeper potential wells); increasing $\tilde{M}_{\mathrm{h}}$ from $10^3-10^4$ suppresses new seed formation at $z\gtrsim11-25$ by factors of $\sim10$; but no significant suppression is seen at $z\sim7-11$. We now summarize the implications of our gas-based seed models on  black hole growth, and their impact on the observable properties of the SMBH populations at $z\geq7$.

\begin{itemize}
    \item For all of our gas-based seed models, the majority of the black holes in our zoom region are primarily growing via black hole mergers. The contribution from gas accretion starts to become significant only at lower redshifts~($z\lesssim1$), particularly for $\gtrsim10^6~M_{\odot}/h$ black holes. That being said, the dominance of merger driven growth at $z>7$ may not persist around more extreme overdense regions; this will be investigated in future work.

    \item Lower seed masses produce dramatically higher black hole merger rates in our models  at high redshift (by factors of $\sim 10$ and $\sim 10^3$ at $z=7$ and $z=15$, respectively, as $M_{\rm seed}$ is decreased from $10^5 M_{\odot}/h$ to $1.56\times10^3 M_{\odot}/h$). This is driven in large part by the fact that the halo and gas mass thresholds are scaled to the seed mass in our models. 
    When we consider only mergers between black holes above a fixed mass threshold that is $\gtrsim 2$ times larger than the seed mass, the impact of the seed mass on the merger rate is much smaller (up to factors of $\sim2-3$). This is not surprising, given that the overall black hole counts above a fixed mass also vary by a similar amount. 

    \item At fixed seed mass, $\tilde{M}_{\mathrm{sf,mp}}$ and $\tilde{M}_{\mathrm{h}}$ have distinct effects on the merger rates. $\tilde{M}_h$ is the most important parameter at the highest redshifts ($z\gtrsim15$), where increasing $\tilde{M}_{\mathrm{h}}$ from $10^3$ to $10^4$ suppress the merger rates by factors of $\gtrsim 100$. At lower redshift, the star-forming, metal-poor gas mass criterion begins to have a larger influence  seed formation. Specifically, an increase in $\tilde{M}_{\mathrm{sf,mp}}$ from 5-150 uniformly suppresses the overall merger rates by factors $\sim8$ over $z\sim7-15$ for $M_{\mathrm{seed}}=1.25\times10^{4}~M_{\odot}/h$.     
    These findings suggest that the redshift distribution of black hole mergers probed by LISA may contain distinct signatures of the underlying seed formation channels.

    \item  The black hole mass functions for different $M_{\mathrm{seed}}$ are similar for black hole masses $\gtrsim10^6~M_{\odot}/h$. This is because lower mass seeds form earlier, which allows them enough time to grow and catch up with higher mass seeds that form later. When the seed mass is fixed, increasing $\tilde{M}_{\mathrm{sf,mp}}$ and $\tilde{M}_{\mathrm{h}}$ substantially suppresses the abundances of $\gtrsim10^6~M_{\odot}/h$. For $M_{\mathrm{seed}}=1.25\times10^4~M_{\odot}/h$, the masses of the heaviest black holes are suppressed by factors of $\sim6$ when $\tilde{M}_{\mathrm{sf,mp}}$ is varied from $5-150$; the suppression is by factors of $\sim10$ when $\tilde{M}_{\mathrm{h}}$ is varied from $10^3-10^4$. This is driven by the fact that black hole growth is completely dominated by mergers; in particular, fewer seeds are formed at higher values of $\tilde{M}_{\mathrm{sf,mp}}$ and $\tilde{M}_{\mathrm{h}}$ resulting in fewer mergers to facilitate black hole growth. 
 
    \item AGN luminosity functions are less sensitive to variations in the seed model parameters; they are suppressed by up to factors $\sim2-3$ with respect to $\tilde{M}_{\mathrm{sf,mp}}$~(from 5-150) and  $\tilde{M}_{\mathrm{h}}$~(from $10^3-10^4$). An increase in $M_{\mathrm{seed}}$ from $1.56\times10^3-10^5~M_{\odot}/h$ does not appreciably change our luminosity functions compared to our statistical uncertainties. We conclude that the overall variation we see in the AGN luminosity functions amongst our seed models will be difficult to observe, even with sensitive future telescopes such as Lynx and JWST. 
\end{itemize}

We must acknowledge that our analysis does not, by any means, encompass the complete parameter space of all possible seed models. Additionally, our limited volume does not allow us to probe the regime of the observed $z>6$ quasars as well as merger rates of massive~($>10^{6}~M_{\odot}/h$) black holes. Additionally, as with all zoom simulations, we are unable to provide volume independent statistics of the SMBH population~(we shall address this using uniform volume simulations in our follow up paper). These factors overall make it difficult for us to directly comment on whether the seed models considered here would be distinguishable with EM or GW observables. Nevertheless, our results do clearly demonstrate that each of the seed model parameters considered here has a large impact on the black hole merger rate detectable with LISA. In contrast, the impact of varying seed parameters on the resulting AGN luminosity functions is relatively modest, indicating that it will likely be difficult to distinguish between these seeding models using high-redshift AGN surveys, even with future instruments such as Lynx. 

Moreover, our results are going to be valuable for continued research because a similar exploration of the parameter space using uniform volume simulations~(at volumes and resolutions similar to IllustrisTNG) will remain infeasible in the near future. To that end, when we do have predictions from a select few uniform volume simulations~(with a specific choice of seed parameters), they can be combined with the results of this work to also make predictions for model variants~(in a cost efficient manner).
 
Recall again that our numerical seed models are 
designed to be agnostic about which physical seed mechanism it is meant to represent. To that end, different values of the seed parameters can be representative of different physical seed formation scenarios. For example, $1.56\times10^{3}~M_{\odot}/h$ seeds forming in $\sim10^6-10^{7}~M_{\odot}/h$ halos may better represent Pop III or NSC seeds~(compared to DCBH seeds). On the other hand, $1\times10^{5}~M_{\odot}/h$ seeds forming in $\sim10^8-10^{9}~M_{\odot}/h$ halos may better represent DCBHs. 
In future work, we plan to explore additional possible seeding criteria such as Lyman Werner intensity from nearby young stars, as well as the infall rates of gas towards halo centers. Finally, this work will drive continued development of gas-based seed models for large volume cosmological simulations. In particular, we plan to use these results for calibrating seed models for uniform volume~(lower resolution) simulations. These results will also be useful for developing new SAMs. In particular, one can apply similar seed models to existing SAMs and compare the results to our work. In the process, one can test the various assumptions that are typically made in SAMs and may bring to light the ones that are most important in the context of black hole seed model development. 

\appendix
\section{Suppressing spurious seeds}
\label{Suppressing short lived seeds and spurious reseeds}

As mentioned in Section \ref{Preventing spurious seeds and mergers}, black holes in our simulation are repositioned to the local potential minima within their neighbor search radii. This may cause some halos to spuriously lose black holes during fly-by encounters around much larger halos, thereby causing these halos to spuriously seed a new BH. Here we describe the implementation for preventing spurious seeding in halos that have experienced a prior seeding event. At the time of new seed formation in a halo, the neighboring gas cells are ``tagged". More specifically, for every gas cell, we create a field called $M^{\mathrm{gas}}_{\mathrm{tag}}$ and set it to be
\begin{equation}
M^{\mathrm{gas}}_{\mathrm{tag}}=M_{\mathrm{seed}}/N_{\mathrm{ngb}}
\end{equation}
where $N_{\mathrm{ngb}}$ is the number of neighboring gas cells around the black hole seed. $M^{\mathrm{gas}}_{\mathrm{tag}}$ is conserved during the refinement and derefinement of gas cells. It is also conserved when the gas cells are converted to star particles; i.e., for a newly formed star particle that was converted from a parent gas cell with $M^{\mathrm{gas}}_{\mathrm{tag}}$, we create a field called $M^{\mathrm{star}}_{\mathrm{tag}}$ and assign 
\begin{equation}
M^{\mathrm{star}}_{\mathrm{tag}} \equiv M^{\mathrm{gas}}_{\mathrm{tag}}.
\end{equation} Next, we compute the total amount of tagged gas and stars for every FOF group by defining $M^{\mathrm{FOF}}_{\mathrm{tag}}$ as
\begin{equation}
M^{\mathrm{FOF}}_{\mathrm{tag}}  \equiv \sum_{\mathrm{gas}}M^{\mathrm{gas}}_{\mathrm{tag}}+\sum_{\mathrm{stars}}M^{\mathrm{stars}}_{\mathrm{tag}}
\end{equation}
where $\sum_{\mathrm{gas}}$ and $\sum_{\mathrm{stars}}$ represent a summation over all the gas cells and star particles respectively within the FOF group.
Finally, for every FOF group we define a quantity $f_{\mathrm{tagged}}$ as
\begin{equation}
f_{\mathrm{tagged}}\equiv M^{\mathrm{FOF}}_{\mathrm{tag}}/M_{\mathrm{seed}}.
\end{equation}

Now, $f_{\mathrm{tagged}}$ is constructed such that if a halo forms a black hole seed and does not lose or gain any tagged gas and stars during its subsequent evolution, then $f_{\mathrm{tagged}}=1$. However, a halo can potentially lose some of its tagged gas and stars due to the usual dynamical processes. Some of the gas can also be swallowed by the BH. Finally, a halo can also acquire tagged gas and stars from other halos. This overall implies that halos can have $f_{\mathrm{tagged}}<1$ or $f_{\mathrm{tagged}}>1$ depending on their assembly history. But the most important point is that the higher the value of $f_{\mathrm{tagged}}$, the higher is the likelihood that the halo had undergone seed formation in the past. We therefore seek to suppress spurious seeding by setting a maximum value of $f_{\mathrm{tagged}}$, above which the halos are prevented from seeding.

\begin{figure*}
\includegraphics[width=12cm]{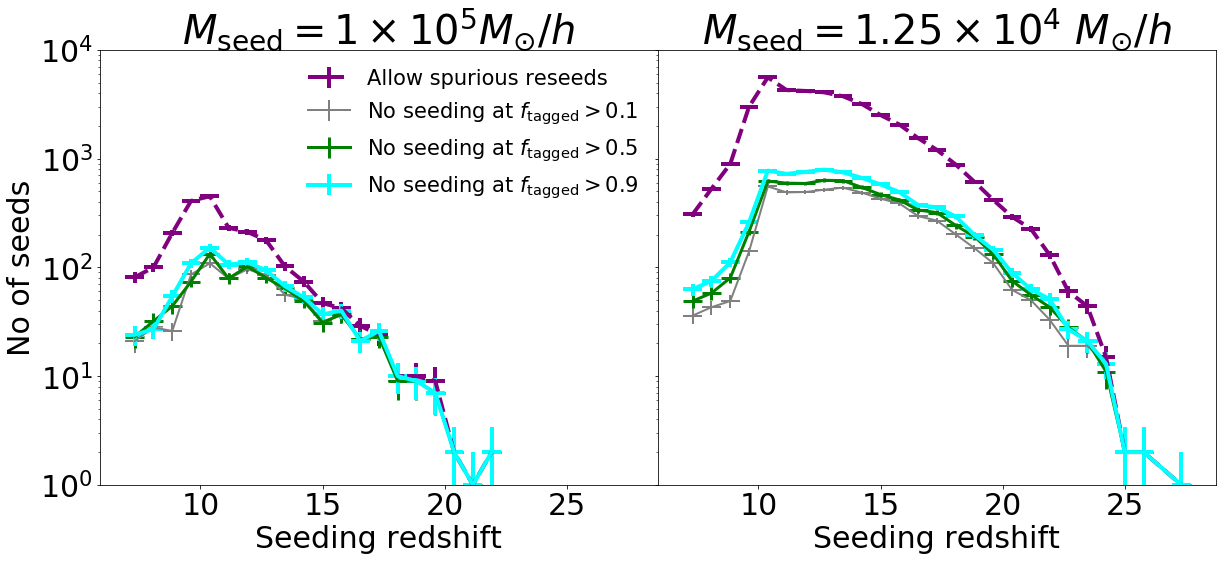} 
\caption{Distribution of seeding redshifts and how it is impacted by the various schemes detailed in Appendix \ref{Suppressing short lived seeds and spurious reseeds} and Section \ref{Preventing spurious seeds and mergers}. These runs were performed with $\tilde{M}_{h}=10^3$, $\tilde{M}_{\mathrm{sf,mp}}=5$~(for \texttt{ZOOM_REGION_z5}). The solid lines correspond to cases where we have suppressed `seeding' in halos that have a ``memory'' of a past seeding event. $f_{\mathrm{tagged}}$ for a halo~(see Appendix \ref{Suppressing short lived seeds and spurious reseeds} for details) is defined such that higher its value, higher is the likelihood that the halo hosted a black hole seed in the past. In particular, black hole seeding is suppressed in halos where $f_{\mathrm{tagged}}$ exceeds a maximum threshold. The cyan, green and grey lines correspond to runs where seeds were only allowed at $f_{\mathrm{tagged}}<0.1,0.5,0.9$ respectively. Purple dashed line corresponds to a run where this ``spurious seeding" is allowed. The difference between the purple dashed line and solid lines show that if we do not suppress the spurious seeds, it leads to substantially higher~(by factors up to 10) number of seeding events. Additionally, we find that after the initial suppression compared to the purple dashed line, the distributions~(cyan, purple and green lines) do not vary significantly with the specific maximum threshold chosen for $f_{\mathrm{tagged}}$ between $0.1-0.9$. We therefore simply choose a fiducial maximum value of $0.5$ for all our runs. Lastly, the prevalence of spurious seeds decreases with increasing $M_{\mathrm{seed}}$}.
\label{Impact_of_seed_suppression_fig}

\includegraphics[width=12cm]{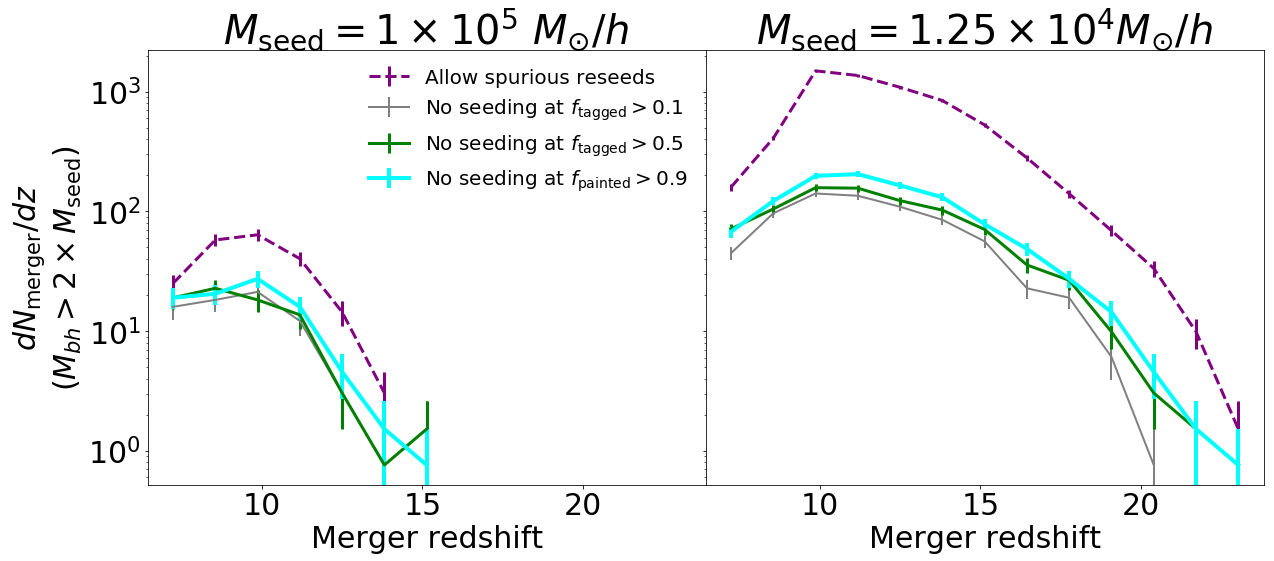} 
\caption{Distribution of merger rates and how it is impacted by the various schemes detailed in Appendix \ref{Suppressing short lived seeds and spurious reseeds} and Section \ref{Preventing spurious seeds and mergers}. These runs were performed with $\tilde{M}_{h}=10^3$, $\tilde{M}_{\mathrm{sf,mp}}=5$~(for \texttt{ZOOM_REGION_z5}). Similar to Figure \ref{Impact_of_seed_suppression_fig}, solid lines correspond to cases where we have suppressed `seeding' in halos that have a ``memory'' of a past seeding event. Cyan, green and grey lines correspond to $f_{\mathrm{tagged}}<0.9,0.5,0.1$. Purple dashed line corresponds to a run where this spurious seeding is allowed. Here also, we see that if the spurious seeds are not suppressed, the merger rates are overestimated by up to factors $\sim10$. Additionally, we find that after the initial suppression compared to the purple dashed line, the merger rates~(cyan, purple and green lines) do not vary significantly with the specific maximum threshold chosen for $f_{\mathrm{tagged}}$ between $0.1-0.9$.}
\label{Impact_of_seed_suppression_merger_rates_fig}
\end{figure*}

In Figures \ref{Impact_of_seed_suppression_fig} and \ref{Impact_of_seed_suppression_merger_rates_fig}, we tested the impact of suppressing the spurious seeds with maximum $f_{\mathrm{tagged}}$ values of 0.1,0.5,0.9. Figure \ref{Impact_of_seed_suppression_fig} shows the distribution of seeding redshifts. We find that our prescription reduces the number of seeding events~(purple dashed lines vs. grey, cyan and green solid lines) by factors up to $\sim10$ for $M_{\mathrm{seed}}=1.25\times10^{4}~M_{\odot}/h$, particularly at $z<15$. Figure \ref{Impact_of_seed_suppression_merger_rates_fig} shows the merger rates, where we see that our prescription suppresses the number of mergers by factors $\sim10$ for $M_{\mathrm{seed}}=1.25\times10^{4}~M_{\odot}/h$. This implies that spurious seeding~(if not suppressed) can substantially contaminate the SMBH populations. As we increase the seed masses~(left vs right panels), spurious seeding is less prevalent. This is expected since higher mass seeds occur in more massive halos, which are less likely to lose their black holes during fly-by encounters. 

Lastly, we also find that after the initial suppression~(purple dashed lines vs. grey, cyan and green solid lines in Figures \ref{Impact_of_seed_suppression_fig} and \ref{Impact_of_seed_suppression_merger_rates_fig}), further changes in the maximum $f_{\mathrm{tagged}}$ threshold from 0.1-0.9~(grey vs cyan vs green lines in \ref{Impact_of_seed_suppression_fig} and \ref{Impact_of_seed_suppression_merger_rates_fig}) do not substantially impact the number of seeds and mergers. This tells us that most halos that had past seeding events do not lose a substantial amount of their tagged gas and stars~(and have $f_{\mathrm{tagged}}>0.9$); this further certifies $f_{\mathrm{tagged}}$ as a very robust diagnostic parameter for identifying and suppressing spurious seeding events. For all the models presented in the remainder of the paper, we choose of maximum threshold of $f_{\mathrm{tagged}}<0.5$ to allow the formation of black hole seeds.

\section*{Acknowledgements}
LB~acknowledges support from National Science Foundation grant AST-1715413. LB and PT acknowledges support from NSF grant AST-1909933 and NASA ATP Grant 80NSSC20K0502. PT also acknowledges support from AST-200849. DN acknowledges funding from the Deutsche Forschungsgemeinschaft (DFG) through an Emmy Noether Research Group (grant number NE 2441/1-1). MV acknowledges support through NASA ATP grants 16-ATP16-0167, 19-ATP19-0019, 19-ATP19-0020, 19-ATP19-0167, and NSF grants AST-1814053, AST-1814259,  AST-1909831 and AST-2007355.

\section*{Data availablity}
The underlying data used in this work shall be made available upon reasonable request to the corresponding author.

\bibliography{references}
\end{document}